\documentclass[journal]{IEEEtran}

\usepackage{amsmath}
\usepackage{amssymb,mathrsfs} 
\usepackage{graphicx}
\usepackage{amsthm}

\usepackage{subfig}
\usepackage[hidelinks]{hyperref} 

\usepackage{bm}
\usepackage{bbm}
\usepackage{amssymb}  
\usepackage{pifont}
\usepackage{array} 
\usepackage{tabularx} 
\usepackage{epstopdf}
\usepackage{dsfont}
\usepackage{color}
\usepackage{epsfig}
\usepackage{amsthm}
\makeatletter

\newcommand{\Rmnum}[1]{\expandafter\@slowromancap\romannumeral #1@}
\makeatother

\usepackage{balance}
\usepackage[utf8]{inputenc}
\usepackage[english]{babel}

\usepackage[mathscr]{euscript}
\usepackage{blkarray,booktabs, bigstrut}
\usepackage{arydshln}

\newtheorem{theorem}{Theorem}
\newtheorem{lemma}{Lemma}
\newtheorem{proposition}{Proposition}

\addto\captionsenglish{\renewcommand{\figurename}{Fig.}}
\usepackage[labelsep=period]{caption}

\usepackage{algorithm}
\usepackage{algorithmicx}
\usepackage{algpseudocode}

\makeatletter
\newcommand{\multiline}[1]{%
  \begin{tabularx}{\dimexpr0.9\linewidth-\ALG@thistlm}[t]{@{}X@{}}
    #1
  \end{tabularx}
}
\makeatother

\begin{document}

\setlength\unitlength{1mm}

\newcommand{\insertfig}[3]{
\begin{figure}[htbp]\begin{center}\begin{picture}(120,90)
\put(0,-5){\includegraphics[width=12cm,height=9cm,clip=]{#1.eps}}\end{picture}\end{center}
\caption{#2}\label{#3}\end{figure}}

\newcommand{\insertxfig}[4]{
\begin{figure}[htbp]
\begin{center}
\leavevmode \centerline{\resizebox{#4\textwidth}{!}{\input
#1.pstex_t}}
\caption{#2} \label{#3}
\end{center}
\end{figure}}

\long\def\comment#1{}


\newfont{\bbb}{msbm10 scaled 700}
\newcommand{\CCC}{\mbox{\bbb C}}

\newcommand{\CC}{\mathbb{ C}}
\renewcommand{\SS}{\mathbb{ S}}
\newcommand{\RR}{\mathbb{ R}}
\newcommand{\PP}{\mathbb{ P}}
\newcommand{\QQ}{\mathbb{ Q}}
\newcommand{\ZZ}{\mathbb{ Z}}
\newcommand{\FF}{\mathbb{ F}}
\newcommand{\GG}{\mathbb{ G}}
\newcommand{\EE}{\mathbb{ E}}
\newcommand{\NN}{\mathbb{ N}}
\newcommand{\KK}{\mathbb{ K}}


\newcommand{\av}{{\bf a}}
\newcommand{\bv}{{\bf b}}
\newcommand{\cv}{{\bf c}}
\newcommand{\dv}{{\bf d}}
\newcommand{\ev}{{\bf e}}
\newcommand{\fv}{{\bf f}}
\newcommand{\gv}{{\bf g}}
\newcommand{\hv}{{\bf h}}
\newcommand{\iv}{{\bf i}}
\newcommand{\jv}{{\bf j}}
\newcommand{\kv}{{\bf k}}
\newcommand{\lv}{{\bf l}}
\newcommand{\mv}{{\bf m}}
\newcommand{\nv}{{\bf n}}
\newcommand{\ov}{{\bf o}}
\newcommand{\pv}{{\bf p}}
\newcommand{\qv}{{\bf q}}
\newcommand{\rv}{{\bf r}}
\newcommand{\sv}{{\bf s}}
\newcommand{\tv}{{\bf t}}
\newcommand{\uv}{{\bf u}}
\newcommand{\wv}{{\bf w}}
\newcommand{\vv}{{\bf v}}
\newcommand{\xv}{{\bf x}}
\newcommand{\yv}{{\bf y}}
\newcommand{\zv}{{\bf z}}
\newcommand{\ellv}{{\bf \ell}}
\newcommand{\zerov}{{\bf 0}}
\newcommand{\onev}{{\bf 1}}


\newcommand{\Am}{{\bf A}}
\newcommand{\Bm}{{\bf B}}
\newcommand{\Cm}{{\bf C}}
\newcommand{\Dm}{{\bf D}}
\newcommand{\Em}{{\bf E}}
\newcommand{\Fm}{{\bf F}}
\newcommand{\Gm}{{\bf G}}
\newcommand{\Hm}{{\bf H}}
\newcommand{\Id}{{\bf I}}
\newcommand{\Jm}{{\bf J}}
\newcommand{\Km}{{\bf K}}
\newcommand{\Lm}{{\bf L}}
\newcommand{\Mm}{{\bf M}}
\newcommand{\Nm}{{\bf N}}
\newcommand{\Om}{{\bf O}}
\newcommand{\Pm}{{\bf P}}
\newcommand{\Qm}{{\bf Q}}
\newcommand{\Rm}{{\bf R}}
\newcommand{\Sm}{{\bf S}}
\newcommand{\Tm}{{\bf T}}
\newcommand{\Um}{{\bf U}}
\newcommand{\Wm}{{\bf W}}
\newcommand{\Vm}{{\bf V}}
\newcommand{\Xm}{{\bf X}}
\newcommand{\Ym}{{\bf Y}}
\newcommand{\Zm}{{\bf Z}}


\newcommand{\At}{\text{A}}
\newcommand{\Bt}{\text{B}}
\newcommand{\Ct}{\text{C}}
\newcommand{\Dt}{\text{D}}
\newcommand{\Et}{\text{E}}
\newcommand{\Ft}{\text{F}}
\newcommand{\Gt}{\text{G}}
\newcommand{\Ht}{\text{H}}
\newcommand{\It}{\text{I}}
\newcommand{\Jt}{\text{L}}
\newcommand{\Kt}{\text{K}}
\newcommand{\Lt}{\text{L}}
\newcommand{\Mt}{\text{M}}
\newcommand{\Nt}{\text{N}}
\newcommand{\Ot}{\text{O}}
\newcommand{\Pt}{\text{P}}
\newcommand{\Qt}{\text{Q}}
\newcommand{\Rt}{\text{R}}
\newcommand{\St}{\text{S}}
\newcommand{\Tt}{\text{T}}
\newcommand{\Ut}{\text{U}}
\newcommand{\Wt}{\text{W}}
\newcommand{\Vt}{\text{V}}
\newcommand{\Xt}{\text{X}}
\newcommand{\Yt}{\text{Y}}
\newcommand{\Zt}{\text{Z}}

\newcommand{\at}{\text{a}}
\newcommand{\bt}{\text{b}}
\newcommand{\ct}{\text{c}}
\newcommand{\dt}{\text{d}}
\newcommand{\et}{\text{e}}
\newcommand{\ft}{\text{f}}
\newcommand{\gt}{\text{g}}
\newcommand{\htx}{\text{h}}
\newcommand{\itx}{\text{i}}
\newcommand{\jt}{\text{j}}
\newcommand{\kt}{\text{k}}
\newcommand{\lt}{\text{l}}
\newcommand{\mt}{\text{m}}
\newcommand{\nt}{\text{n}}
\newcommand{\ot}{\text{o}}
\newcommand{\pt}{\text{p}}
\newcommand{\qt}{\text{q}}
\newcommand{\rt}{\text{r}}
\newcommand{\st}{\text{s}}
\newcommand{\ttx}{\text{t}}
\newcommand{\ut}{\text{u}}
\newcommand{\vt}{\text{v}}
\newcommand{\wt}{\text{w}}
\newcommand{\xt}{\text{x}}
\newcommand{\yt}{\text{y}}
\newcommand{\zt}{\text{z}}


\newcommand{\Ac}{{\cal A}}
\newcommand{\Bc}{{\cal B}}
\newcommand{\Cc}{{\cal C}}
\newcommand{\Dc}{{\cal D}}
\newcommand{\Ec}{{\cal E}}
\newcommand{\Fc}{{\cal F}}
\newcommand{\Gc}{{\cal G}}
\newcommand{\Hc}{{\cal H}}
\newcommand{\Ic}{{\cal I}}
\newcommand{\Jc}{{\cal J}}
\newcommand{\Kc}{{\cal K}}
\newcommand{\Lc}{{\cal L}}
\newcommand{\Mc}{{\cal M}}
\newcommand{\Nc}{{\cal N}}
\newcommand{\Oc}{{\cal O}}
\newcommand{\Pc}{{\cal P}}
\newcommand{\Qc}{{\cal Q}}
\newcommand{\Rc}{{\cal R}}
\newcommand{\Sc}{{\cal S}}
\newcommand{\Tc}{{\cal T}}
\newcommand{\Uc}{{\cal U}}
\newcommand{\Wc}{{\cal W}}
\newcommand{\Vc}{{\cal V}}
\newcommand{\Xc}{{\cal X}}
\newcommand{\Yc}{{\cal Y}}
\newcommand{\Zc}{{\cal Z}}


\newcommand{\RNum}[1]{\uppercase\expandafter{\romannumeral #1\relax}}


\newcommand{\alphav}{\hbox{\boldmath$\alpha$}}
\newcommand{\betav}{\hbox{\boldmath$\beta$}}
\newcommand{\gammav}{\hbox{\boldmath$\gamma$}}
\newcommand{\deltav}{\hbox{\boldmath$\delta$}}
\newcommand{\etav}{\hbox{\boldmath$\eta$}}
\newcommand{\lambdav}{\hbox{\boldmath$\lambda$}}
\newcommand{\epsilonv}{\hbox{\boldmath$\epsilon$}}
\newcommand{\nuv}{\hbox{\boldmath$\nu$}}
\newcommand{\muv}{\hbox{\boldmath$\mu$}}
\newcommand{\zetav}{\hbox{\boldmath$\zeta$}}
\newcommand{\phiv}{\hbox{\boldmath$\phi$}}
\newcommand{\psiv}{\hbox{\boldmath$\psi$}}
\newcommand{\thetav}{\hbox{\boldmath$\theta$}}
\newcommand{\tauv}{\hbox{\boldmath$\tau$}}
\newcommand{\omegav}{\hbox{\boldmath$\omega$}}
\newcommand{\xiv}{\hbox{\boldmath$\xi$}}
\newcommand{\sigmav}{\hbox{\boldmath$\sigma$}}
\newcommand{\piv}{\hbox{\boldmath$\pi$}}
\newcommand{\rhov}{\hbox{\boldmath$\rho$}}

\newcommand{\Gammam}{\hbox{\boldmath$\Gamma$}}
\newcommand{\Lambdam}{\hbox{\boldmath$\Lambda$}}
\newcommand{\Deltam}{\hbox{\boldmath$\Delta$}}
\newcommand{\Sigmam}{\hbox{\boldmath$\Sigma$}}
\newcommand{\Phim}{\hbox{\boldmath$\Phi$}}
\newcommand{\Pim}{\hbox{\boldmath$\Pi$}}
\newcommand{\Psim}{\hbox{\boldmath$\Psi$}}
\newcommand{\Thetam}{\hbox{\boldmath$\Theta$}}
\newcommand{\Omegam}{\hbox{\boldmath$\Omega$}}
\newcommand{\Xim}{\hbox{\boldmath$\Xi$}}

\newcommand{\supp}{{\hbox{supp}}}
\newcommand{\sinc}{{\hbox{sinc}}}
\newcommand{\diag}{{\hbox{diag}}}
\renewcommand{\det}{{\hbox{det}}}
\newcommand{\trace}{{\hbox{tr}}}
\newcommand{\sign}{{\hbox{sign}}}
\renewcommand{\arg}{{\hbox{arg}}}
\newcommand{\var}{{\hbox{var}}}
\newcommand{\cov}{{\hbox{cov}}}
\newcommand{\SINR}{{\sf SINR}}
\newcommand{\SNR}{{\sf SNR}}
\newcommand{\Ei}{{\rm E}_{\rm i}}
\renewcommand{\Re}{{\rm Re}}
\renewcommand{\Im}{{\rm Im}}
\newcommand{\eqdef}{\stackrel{\Delta}{=}}
\newcommand{\defines}{{\,\,\stackrel{\scriptscriptstyle \bigtriangleup}{=}\,\,}}
\newcommand{\<}{\left\langle}
\renewcommand{\>}{\right\rangle}
\newcommand{\herm}{{\sf H}}
\newcommand{\transp}{{\sf T}}
\renewcommand{\vec}{{\rm vec}}


\newcommand{\GameNF}{\mathcal{G} = \left(\mathcal{K}, \left\lbrace\mathcal{A}_k \right\rbrace_{k \in \mathcal{K}},\phi \right)}
\newcommand{\gameNF}{\mathcal{G}}
\newcommand{\BR}{\mathrm{BR}}


\newcommand{\squeezeequ}{\medmuskip=2mu \thinmuskip=1mu \thickmuskip=3mu}
\newcommand{\middlesqueezeequ}{\medmuskip=1.5mu \thinmuskip=0.5mu \thickmuskip=2.5mu \nulldelimiterspace=-0.5pt \scriptspace=0pt}
\newcommand{\supersqueezeequ}{\medmuskip=1mu \thinmuskip=0mu \thickmuskip=2mu \nulldelimiterspace=-1pt \scriptspace=0pt}
\newcommand{\Dsupersqueezeequ}{\medmuskip=0.5mu \thinmuskip=0mu \thickmuskip=1mu \nulldelimiterspace=-1pt \scriptspace=0pt}
\newcommand{\Tsupersqueezeequ}{\medmuskip=0.1mu \thinmuskip=0mu \thickmuskip=0.1mu \nulldelimiterspace=-1pt \scriptspace=0pt}

\newcommand{\Cov}[1]
{\Sigmam_{\mathbf{#1}}}

\def\LRT#1#2{\!
\raisebox{.2ex}{$
{{\scriptstyle\;#1}\atop{\displaystyle\gtrless}}
\atop
{\raisebox{-1.25ex}{$\scriptstyle\;#2$}}
$}
\!}

\newcommand{\squeezedequation}{\medmuskip=2mu \thinmuskip=1mu \thickmuskip=3mu}
\newcommand{\supersqueezedequation}{\medmuskip=1mu \thinmuskip=0mu \thickmuskip=2mu \nulldelimiterspace=-1pt \scriptspace=0pt}

%
\title{
The Role of Information Incompleteness in Defending Against Stealth Attacks
}
%
%
%
\author{Ke Sun,~\IEEEmembership{Senior~Member,~IEEE,}
   Jingyi Yan, Zhenglin Li, and Shaorong Xie

\thanks{
 \emph{(Corresponding author: Zhenglin Li and Shaorong Xie)}

K. Sun and J. Yan are with the School of Computer Engineering and Science, Shanghai University, Shanghai 200444, China; 
K. Sun is also with Shanghai Key Laboratory of Intelligent Connected Vehicle Cybersecurity, Shanghai 200444, China
(email: ke\_sun@shu.edu.cn, jingyiyan@shu.edu.cn).

Z. Li is with the School of Future Technology, Shanghai University, Shanghai 200444, China 
(email: zhenglin\_li@shu.edu.cn).

S. Xie is with the School of Mechatronic Engineering and Automation, Shanghai University, Shanghai 200444, China (email: srxie@shu.edu.cn).

}}
%



\maketitle

\begin{abstract}
The effectiveness of Data Injection Attacks (DIAs) critically depends on the completeness of the system information accessible to adversaries. 
This relationship positions information incompleteness enhancement as a vital defense strategy for degrading DIA performance.
In this paper, we focus on information-theoretic stealth attacks, where the attacker encounters a fundamental tradeoff between the attack stealthiness and destructiveness. 
Specifically, we systematically characterize how incomplete admittance information impacts these dual objectives. 
In particular, we establish sufficient conditions for two distinct operational regimes: (i) stealthiness intensifies while destructive potential diminishes and (ii) destructiveness increases while stealth capability weakens.
For scenarios beyond these regimes, we propose a maximal incompleteness strategy to optimally degrade stealth capability.
To solve the associated optimization problem, the feasible region is reduced without excluding the optimal
solution, and a heuristic algorithm is then introduced to effectively identify the near-optimal solutions within the reduced region. 
Furthermore, we extend our theoretical analysis to broader scenarios in which the initial sufficient conditions are not met, introducing exact diagnostic indices to evaluate instance-wise performance shifts.
Numerical simulations are conducted on IEEE test systems to validate the findings.
\end{abstract}


%
\IEEEpeerreviewmaketitle

\section{Introduction}
%
%
%
%

\IEEEPARstart{D}{ata} injection attacks (DIAs) are cybersecurity threats that compromise the data integrity of state estimation procedures within the Smart Grid. 
Attack performance is typically evaluated based on two interdependent criteria, attack stealthiness and attack destructiveness.
Specifically, the attacker manipulates measurements (i.e. the data) used for state estimation to disrupt system operations \cite{giani_viking_2009,liu_false_2009}, representing the attack’s {\it destructiveness}. 
Simultaneously, the attacker seeks to evade the bad data detection (BDD) mechanism deployed by the grid operator, i.e. comprising measurements without triggering the BDD, reflecting the attack’s {\it stealthiness}.
To that end, the attacker requires varying levels of system information depending on the state estimation and BDD methods in place.
For example, disrupting weighted least squares (WLS) estimation and bypassing residual-based detection necessitates knowledge of the system Jacobian matrix \cite{liu_false_2009}. 
In the case of minimum mean squared error estimation (MMSE) under a Bayesian framework, the attacker requires both the Jacobian matrix and the second-order statistics of the state variables to optimize the tradeoff between destructiveness and stealthiness \cite{kosut_malicious_2011, esnaola_maximum_2016, Sun_information-theoretic_2017, Sun_Stealth_2020}.

However, the attacker typically has access only to  incomplete system information, which consequently degrades attack performance in terms of stealthiness or destructiveness.
This incompleteness may stem from limited capabilities of the attacker, dedicated defense mechanisms (such as moving target defense (MTD) \cite{MTD, yan_2026_bt}), or imperfect learning of unknown system information \cite{sun_2019_learning, sun_2023_asymptotic}. 
From the perspective of the operator, such information incompleteness can serve as an effective defense against stealth attacks \cite{irfan_2024_survey}.
For example, it is shown in \cite{rahman_false_2012} that an inaccurate system Jacobian matrix indeed increases the probability of detection under residual-based detection. 
Similar findings include \cite{sun_2019_learning, sun_2023_asymptotic, esmalifalak_2011_stealth, deng_2018_false, tajer_false_2019}, and so on. 
Notably, the aforementioned studies mainly focus on the WLS estimation and residual-based detection setting; while \cite{sun_2019_learning} and \cite{sun_2023_asymptotic} extend the analysis to a Bayesian framework, only suboptimality of the attacks is proved.
However, the broader impact of incomplete information on attack performance under a Bayesian framework remains understudied. 
Specifically, incomplete information alters the tradeoff between destructiveness and stealthiness, deviating it from the optimal balance.
The impact of incomplete information on the nature of tradeoff adjustments remains insufficiently explored.

\subsection{Related Works}

Under an incomplete system information setting, attackers aim to construct stealth attacks using various forms of incomplete information. 
Specifically, studies such as \cite{sun_2024_stealth, liu_2025_low} examine scenarios where branch admittance information is partially unknown, while \cite{liu_2025_low, liu_local_2014, liu_false_2017, jin_2024_false} explore cases where only part of the system is fully known. 
The impact of incomplete topology information is analyzed in \cite{qu_2024l_ocalization}, whereas \cite{yu_blind_2015, zhang_2018_can, higgins_2021_topology} investigate scenarios where the incompleteness arises from the learning of unknown system information.
These works focus on the WLS estimation and residual-based detection case, where perfect stealth attacks (i.e.,  attacks that induce no change in residuals) are possible if the attacker has complete information \cite{liu_false_2009}. 

However, under a Bayesian framework, the probability of detection remains nonzero whenever any nonzero attack is launched, which implies that the perfect stealth attacks do not exist even with complete system information \cite{kosut_malicious_2011, esnaola_maximum_2016, Sun_information-theoretic_2017, Sun_Stealth_2020}.
As a result, the attacker must balance attack destructiveness and stealthiness.
In the presence of incomplete information, this tradeoff between destructiveness and stealthiness is altered and deviated from the optimal balance, leading to the suboptimality of the attacks  \cite{sun_2019_learning, sun_2023_asymptotic}. 
However, suboptimality alone only indicates that the tradeoff has changed, it remains unclear how exactly incomplete information affects the tradeoff:  whether destructiveness and stealthiness increase or decrease simultaneously, or if one increases while the other decreases.

Note that although stealth or nearly stealth attacks may still be feasible under incomplete system information, the destructiveness or stealthiness of the attacks is typically degraded \cite{srivastava_2018_graph}. 
The attacker has to decrease the destructiveness (cannot compromise arbitrary state variables or measurements \cite{sun_2024_stealth, liu_local_2014, liu_false_2017}), or acquires additional information (such as historical measurements \cite{sun_2023_asymptotic, liu_2025_low, zhang_2018_can}), to compensate for the lack of necessary information. 
This creates opportunities for operators to detect or mitigate such attacks. 
The interplay between the attacker and the operator under incomplete information setting is studied in \cite{ ghosh_2020_defending, yan_2024_game, yi_2024_defense}, in which a Bayesian game-theoretic model is developed and the corresponding Bayesian equilibrium is obtained.

\subsection{Contributions}

In this paper, we focus on the Bayesian framework and information-theoretic attacks (see Section \ref{Subsec:ITA}), analyzing the impact of incomplete admittance information on performance degradation. 
Specifically, we characterize how the tradeoff between attack stealthiness and destructiveness changes under the incomplete admittance information setting, and propose a strategy to maximize the degradation in stealthiness.

\begin{figure}[t!]
\centering
\includegraphics[scale=0.62]{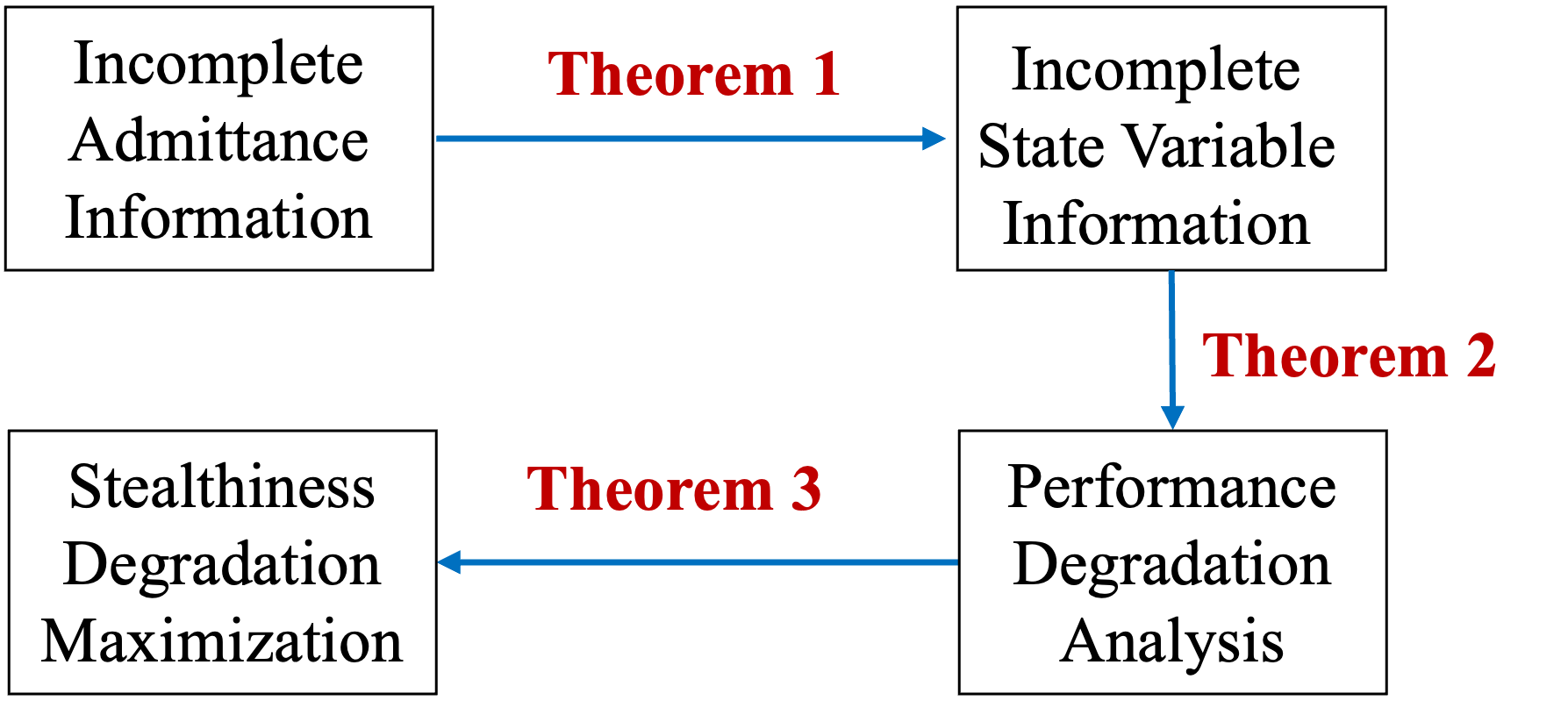}
\caption{Contributions of the paper.}
\label{Fig:MI}
\end{figure}

As illustrated in Fig. \ref{Fig:MI}, firstly, the impact of the incomplete admittance information on the attack construction strategy is characterized, in which the incompleteness of admittance information is reformulated into the incompleteness of state variable information (Theorem \ref{theorem:EquMismatch}). 
Based on this foundation, the attack performance degradation is analyzed, providing sufficient conditions for the regime in which stealthiness intensifies while destructive potential diminishes, and for the case in which destructiveness intensifies while stealth capability weakens (Theorem \ref{Lemma:Delta_Con} and Lemma \ref{Lemma:SC_Delta}). 
For scenarios beyond these regimes, an optimization problem is formulated to maximize stealth degradation. 
To solve this problem efficiently, the feasible region is reduced without excluding the optimal solution (Theorem \ref{theorem:PDMax}), and a heuristic algorithm (Algorithm \ref{Algo:greedy_1}) is then introduced to effectively identify the solution within this reduced region. 
In addition, a practical way for inducing such incomplete admittance information is proposed (Section \ref{Sec:PI}). 
Finally, we extend our theoretical framework to broader scenarios in which the aforementioned sufficient conditions are not satisfied, establishing exact diagnostic indices to assess instance-wise shifts in stealthiness and destructiveness (Section \ref{Sec:FE}).

The remainder of the paper is organized as follows. 
Section \ref{Sec:System_Model} introduces the system model. 
Section \ref{Sec:ProbForm} analyzes the impact of the incomplete admittance information on the attack performance, in which the incompleteness reformulation result is presented. 
Section \ref{Sec:IAIIPD} examines the resulting performance degradation, in which the sufficient conditions, the maximal stealthiness degradation strategy, and the practical implementations are proposed. 
Section \ref{Sec:FE} provides the exact diagnostic indices to evaluate performance in broader scenarios.
The findings in Section \ref{Sec:ProbForm} and \ref{Sec:IAIIPD} are numerically evaluated in Section \ref{Sec:Sim}. 
Finally, Section \ref{Sec:Con} concludes the paper.

\section{System Model} \label{Sec:System_Model}
The notation used in this paper is described below.  
Vectors of random variables are represented by capital letters with a superscript indicating the dimension, such as $X^n$. 
Sets are represented using calligraphy font, such as $\Rc$ denoting the set of real numbers, $\Sc^{m}_{+}$ denoting the set of positive semi-definite (PSD) matrices of size $m\times m$, $\Sc^{m}_{++}$ denoting the set of positive definite (PD) matrices of size $m\times m$, and $\Dc^{m}$ denoting the set of diagonal matrices of size $m\times m$. 
Matrices are represented by bold capital letters, such as $\Am$, and the entry in the $i$-th row and $j$-th column of matrix $\Am$ is represented by $A_{i,j}$.
The matrix that is formed by selecting a given set of columns from $\Am$ is represented by $\Am_{\cdot, \mathcal{I}}$, where $\mathcal{I}$ is the set of the column indices. 
The identity matrix of proper dimensions is denoted by $\Id$, and the matrix or vector whose elements are all zero is denoted by $\zerov$.
Vectors are represented by boldface lowercase letters, such as $\av$, whose $\ell_2$ norm is denoted by $\|\av\|_2$. 
$\textnormal{diag}\left(a_1, \ldots,  a_n \right)$ denotes a diagonal matrix with diagonal elements $a_1, \ldots,  a_n$. 
$\Am \geq \Bm$ ($\Am \leq \Bm$) refers to elementwise greater than or equal to (less than or equal to), which also applies to scalars and vectors. 
$\Am \preceq \Bm$ ($ \Am \succeq \Bm$) is defined in terms of Loewner order of symmetric matrices \cite[pp.219]{seber_matrix_2008}.

\subsection{Linearized Observation Model}
The measurement model for state estimation under linearized dynamics is given by
\begin{IEEEeqnarray}{c}\label{Equ:DCSE}
Y^{m} = \Hm X^{n} + Z^{m},
\end{IEEEeqnarray}
where $Y^{m} \in \Rc^{m}$ is a vector of random variables describing the measurements; 
$X^{n} \in \Rc^{n}$ is a vector of random variables describing the state variables;
$\Hm \in \Rc^{m\times n}$ is the linearized Jacobian measurement matrix, which is determined by the power network topology and the admittance of the branches;
and $Z^{m} \in \Rc^{m} $ is the additive white Gaussian noise distributed as $\Nc(\zerov,\sigma^{2} \Id)$, in which $\sigma^2$ is the variance of the noise \cite{abur_power_2004, grainger_power_1994}.

In the following, we assume that the vector of state variables follows a zero-mean multivariate Gaussian distribution given by
\begin{IEEEeqnarray}{c}\label{Equ:SV_Gaussian}
  X^{n} \sim \Nc(\zerov,\Sigmam_{X\!X}),
\end{IEEEeqnarray}
where $\Sigmam_{X\!X}\in \Sc^{n}_{+}$ is the covariance matrix of the state variables, and the rationale for this modeling assumption is discussed in \cite[Section 3.1.1]{sun_thesis}.
Due to the linearized model in (\ref{Equ:DCSE}), the vector of measurements also follows a multivariate Gaussian distribution denoted by
\begin{IEEEeqnarray}{c}\label{Equ:measurement_dis}
  Y^{m} \sim \Nc(\zerov,\Sigmam_{Y\!Y}),
\end{IEEEeqnarray}
where $\Sigmam_{Y\!Y} = \Hm\Sigmam_{X\!X}\Hm^{\sf T} + \sigma^{2}\Id$ is the corresponding covariance matrix.

\subsection{Property of Jacobian Matrix}

The structure of the Jacobian matrix $\Hm$ is determined by the topology of the system and the admittance of the branches. 
Let $n + 1$ and $l$ denote the number of buses and branches, respectively. 
For each $i \in \{ 1, \dots, l \}$, let $\mathsf{e}_i$ represent the $i$-th branch and $b_i$ represent the corresponding admittance of $\mathsf{e}_i$. 
And let ${\cal E} = \{\mathsf{e}_1, \ldots, \mathsf{e}_l\}$ denote the set of all branches.
Hence, the diagonal branch admittance matrix is given by 
\begin{IEEEeqnarray}{c}\label{Equ:D}
\Dm \eqdef \textnormal{diag}\left(b_1, \ldots,  b_l \right); 
\end{IEEEeqnarray}
and the branch-bus incidence matrix is denoted by $\bar{\Am} \in \{ -1, 0, 1 \}^{l \times \left( n+1 \right)}$, whose elements satisfy
\begin{IEEEeqnarray}{c}
\bar{A}_{ki} = \left\{ \,
\begin{IEEEeqnarraybox}[][c]{l?s}
\IEEEstrut
1, & if branch $\mathsf{e}_k$ starts from bus $i$, \\
-1, & if branch $\mathsf{e}_k$ ends at bus $i$, \\
0, & otherwise.
\IEEEstrut
\end{IEEEeqnarraybox}
\right.
\label{eq:example_left_right1}
\end{IEEEeqnarray}
Note that if the $i$-th bus is selected as the reference bus, the $i$-th column needs to be removed from $\bar{\Am}$.
Following the setting of \cite[pp. 259]{grainger_power_1994}, we select the first bus as the reference bus, which brings about a reduced branch-bus incidence matrix $\Am \in \{ -1, 0, 1 \}^{l \times  n }$ and 
\begin{IEEEeqnarray}{c}\label{Equ:BBIM}
\Am \eqdef \bar{\Am}_{\cdot, \{2, \ldots, n+1 \}}.
\end{IEEEeqnarray}

Choosing the net active power injections of buses and the active power flow in both directions as measurements, and the bus voltage angle difference with respect to the reference bus as state variables, yields the following Jacobian matrix. 
\begin{IEEEeqnarray}{c}\label{Equ:H}
\Hm
    = 
    \begin{bmatrix}
    \Am^{\sf T} \Dm \Am \\
    \Dm \Am \\
    - \Dm \Am
    \end{bmatrix}
= 
    \begin{bmatrix}
    \Am^{\sf T}  \\
     \Id\\
    - \Id
    \end{bmatrix}\Dm \Am
 \eqdef \Jm \Dm \Am,     
\end{IEEEeqnarray}
in which $\Jm \eqdef \left[ \Am, \Id, - \Id\right]^{\sf T}$.
Under the assumption that the power grid is a connected graph (with branches as edges and buses as vertices) and the matrix $\Dm$ is a full rank matrix, it can be known that matrix $\Hm$ is a full rank matrix \cite[Lemma 1 and (14)]{sun_2024_stealth}.

\subsection{Attack and Detection Model within Bayesian Framework}
DIAs compromise the measurements available to the operator by adding an attack vector of random variables to the measurements.
The resulting vector of compromised measurements is given by
\begin{IEEEeqnarray}{c}
\label{eq:measurement_model}
Y^{m}_{A} = \Hm X^{n} + Z^{m} + A^{m},
\end{IEEEeqnarray}
where $A^{m} \in \Rc^{m}$ is the attack vector and $Y^{m}_{A} \in \Rc^{m } $ is the vector containing the compromised measurements \cite{liu_false_2009}.

Following the setting of \cite{Sun_information-theoretic_2017}, \cite{Sun_Stealth_2020}, and \cite{sun_2023_asymptotic},
an attack vector which is independent of the state variables is constructed under the assumption that the attack vector follows a multivariate Gaussian distribution denoted by
\begin{IEEEeqnarray}{c}\label{Equ:attack}
  A^{m} \sim  \Nc (\zerov,\Sigmam_{A\!A}), 
\end{IEEEeqnarray}
where $\Sigmam_{A\!A}\in \Sc^{m}_{+}$ is the associated covariance matrix.
Due to the Gaussianity of the attack, the vector of compromised measurements is distributed as
\begin{IEEEeqnarray}{c}\label{Equ:Dis_Ya}
  Y_{A}^{m} \sim \Nc(\zerov,\Sigmam_{Y_{A}\!Y_{A}}),
\end{IEEEeqnarray}
where $\Sigmam_{Y_{A}\!Y_{A}} = \Hm\Sigmam_{X\!X}\Hm^{\sf T} + \sigma^{2}\Id + \Sigmam_{A\!A}$. 

The power system operator makes use of the acquired measurements to detect the attack.
The detection problem is cast as a hypothesis testing problem with hypotheses
\begin{IEEEeqnarray}{cl}
\Hc_{0}:  \ & Y^{m} \sim \Nc(\zerov,\Sigmam_{Y\!Y}), \quad \text{versus} \label{Equ:HT_0} \\
\Hc_{1}:  \ & Y^{m} \sim \Nc(\zerov,\Sigmam_{Y_{A}\!Y_{A}}). \label{Equ:HT_1} 
\end{IEEEeqnarray}
The likelihood ratio test (LRT), which is optimal under the Neyman-Pearson criterion \cite[Proposition \RNum{2}.D.1]{poor_introduction_1994}, is employed to decide between $\Hc_{0}$ and $\Hc_{1}$.
The LRT between $\mathcal{H}_{0}$ and $\mathcal{H}_{1}$ takes the following form:
\begin{IEEEeqnarray}{c}
L ( \yv ) \eqdef \frac{f_{Y^{m}_{A}}(\yv)}{f_{Y^{m}}(\yv)} \ \LRT{\Hc_{1}}{\Hc_{0}} \ \tau, 
\label{LHRT}
\end{IEEEeqnarray}
where $\yv \in \Rc^{m}$ is a realization of the vector of random variables modeling the measurements; $f_{Y_A^m}$ and  $f_{Y^m}$ denote the probability density functions of $Y_A^m$ and  $Y^m$, respectively; and $\tau$ is the decision threshold determined by the false alarm constraint.

\subsection{Information-Theoretic Attacks}\label{Subsec:ITA}
The objective of information-theoretic attacks is two-fold.
Firstly, the attacker seeks to disrupt the information acquisition procedure in (\ref{Equ:DCSE}) by minimizing the amount of information that the operator can acquire from the compromised measurements.
This is accomplished by minimizing the mutual information between the vector of state variables and the vector of compromised measurements, i.e. minimizing $I(X^{n};Y_{A}^{m})$, where $I(\cdot \hspace{0.1em};\cdot)$ denotes the mutual information{\footnote{As shown in \cite{SA_TIT_13}, for the model in \eqref{eq:measurement_model}, the Gaussian distribution minimizes the mutual information between the state variables and the compromised measurements among all additive attack distributions with a prescribed covariance \cite[Section 3.1.2]{sun_thesis}. 
In other words, a Gaussian attack is the least informative from the defender’s perspective under a fixed second-order constraint.
This information-theoretic optimality provides a clear justification for modeling the attack vector in \eqref{Equ:attack} as Gaussian random variables.}}.
Under the Bayesian framework in \eqref{Equ:DCSE}, such a reduction in mutual information directly leads to an increase in the MMSE of the operator’s state estimator, thereby degrading estimation accuracy \cite[Section 3.2.1]{sun_thesis}.

Secondly, the attacker aims to ensure the stealthiness of the attacks under the adopted detection scheme, which involves the minimization of the probability of detection under the LRT in (\ref{LHRT}). 
And as a result of the Chernoff-Stein lemma \cite[Theorem 11.7.3]{cover_elements_2012}, minimizing the probability of detection is equivalent to minimizing the Kullback–Leibler (KL) divergence between the probability distributions of the compromised and uncompromised measurements.
That is, the attacker minimizes $D(P_{Y^{m}_{A}}\|P_{Y^{m}})$, where $P_{Y_A^m}$ and $P_{Y^m}$ denote the probability distributions of $Y_A^m$ and $Y^m$, respectively, and $D(\cdot \|\cdot)$ denotes the Kullback-Leibler (KL) divergence.

The \emph{information-theoretic attacks} minimize the amount of information acquired by the operator and the probability of attack detection simultaneously by minimizing the \textit{data integrity} cost function given by 
\begin{IEEEeqnarray}{c}\label{Equ:Stealth_obj}
\underset{A^{m}}{\text{min}} \ I\left(X^{n};Y^{m}_{A}\right)  +   D\left( P_{{Y}^{m}_{A}}\|P_{Y^{m}} \right).
\end{IEEEeqnarray}
Under the Gaussian assumptions for the state variables and the attack, it is shown in \cite{Sun_information-theoretic_2017, esnaola_data_2021} 
 that the data integrity cost function is a convex function of the attack covariance matrix $\Sigmam_{A\!A}$ in \eqref{Equ:attack} and the optimal solution is given by 
\begin{IEEEeqnarray}{c}\label{Equ:Stealth_optimal}
 \Sigmam_{A\!A}^{\star} = \Hm\Sigmam_{X\!X}\Hm^{\sf T}.
\end{IEEEeqnarray}

\subsection{Incomplete System Information}\label{Subsec:MTD}
The effectiveness of DIAs depends on the amount of system information available to the attacker. 
And in practical scenarios, the attacker may have access only to  incomplete system information, which results in attack performance degradation. 
Specifically, in this paper, {\it we consider the case in which the incompleteness lies in the branch admittance values, i.e. the attacker does not perfectly know the admittance of the branches.}
Furthermore, we assume that the attacker knows the topology of the system, or equivalent matrix $\Am$ in \eqref{Equ:BBIM}, whose rationality lies in the fact that the topology of power system typically change infrequently. 

Let ${\cal E_{I}} \subset \Ec$ be the set of branches whose admittances are not perfectly known by the attacker, and without loss of generality, we assume that ${\cal E_{I}} = \{ \mathsf{e}_1, \ldots, \mathsf{e}_k\}$, i.e. the admittance of the first $k$ branches are not perfectly known by the attacker. 
And let $\Dm'\in \Dc^{l}$ be the diagonal branch admittance matrix constructed from the information available to the attacker. 
Specifically, the matrix $\Dm'$ is given by 
\begin{IEEEeqnarray}{c}\label{Equ:DBAMTD}
\Dm' = \textnormal{diag} \left(b_1', \ldots, b_l' \right),
\end{IEEEeqnarray}
and it holds for all $i \in \{ 1, \dots, l \}$ that 
\begin{IEEEeqnarray}{c}\label{Equ:DBAMTD1}
b_i' = \left\{ 
\begin{tabular}{cl}
$b_i$, & \textnormal{if $\mathsf{e}_i \not\in {\cal E_{I}}$} ,\\
$ b_i + \Delta b_i =  (1+\phi_i)b_i$, & \textnormal{if $\mathsf{e}_i \in {\cal E_{I}}$}, \\
\end{tabular}
\right. 
\end{IEEEeqnarray}
where $b_i$ is the true admittance of branch $\mathsf{e}_i$ defined above \eqref{Equ:D}; $\Delta b_i$ denotes the difference between the admittance known by the attacker and the true admittance; and $\phi_i \eqdef \frac{\Delta b_i}{b_i}$ is the ratio between the difference and the true admittance, which serves as a quantitative measure of the admittance information incompleteness. 

Hence, from \eqref{Equ:H}, the Jacobian matrix $\Hm'$ that is formed by the information the attacker has is given by 
\begin{IEEEeqnarray}{c}
    \Hm'  = 
    \begin{bmatrix}
    \Am^{\sf T} \Dm' \Am \\
    \Dm' \Am \\
    - \Dm' \Am
    \end{bmatrix}
    = 
    \begin{bmatrix}
    \Am^{\sf T} \\
    \Id \\
    - \Id
    \end{bmatrix}
    \Dm' \Am
    =     \Jm \Dm' \Am = \Jm \left(\Id + {\bf \Phi}\right) \Dm \Am ,\label{Equ:NH} \supersqueezeequ \IEEEeqnarraynumspace
\end{IEEEeqnarray}
where $\Jm$ is defined in \eqref{Equ:H}; and ${\bf \Phi} \in \Dc^{l}$ is a diagonal matrix whose diagonal elements are $\phi_i$, i.e. 
\begin{IEEEeqnarray}{c}\label{Equ:alpha}
{\bf \Phi} \eqdef \textnormal{diag} \left(\phi_1, \ldots, \phi_k, 0, \ldots, 0\right)
\end{IEEEeqnarray}
with $\phi_i$ in \eqref{Equ:DBAMTD1}.


\section{Performance of Attacks Constructed via Incomplete Admittance Information}\label{Sec:ProbForm}

As mentioned in the aforementioned text, incomplete system information degrades the attack performance in terms of stealthiness and/or destructiveness.
In the context of information-theoretic attacks, incomplete knowledge of branch admittances compels the attacker to construct an inaccurate Jacobian matrix.  
Consequently, the resulting attack covariance matrix deviates from the optimal one derived under perfect system knowledge, altering the tradeoff between stealthiness and destructiveness. 
To analyze the altered tradeoff, the impact
of incomplete admittance information on the stealthiness and destructiveness needs to be characterized first, which is the focus of this section.

Specifically, as established in Section \ref{Subsec:ITA} and \ref{Subsec:MTD}, the optimal information-theoretic attack under complete system knowledge is given by
\begin{IEEEeqnarray}{c}\label{Equ:A_tilde}
\tilde{A}^m \sim \Nc \left( \zerov, \Hm \Sigmam_{X\!X} \Hm^{\sf T}\right).
\end{IEEEeqnarray}
However, in the presence of incomplete admittance information, the attacker lacks access to the true Jacobian matrix $\Hm$ and instead constructs a Jacobian matrix $\Hm'$ as defined in \eqref{Equ:NH}.
This results in a suboptimal attack
\begin{IEEEeqnarray}{c}\label{Equ:A_bar}
\bar{A}^m \sim \Nc \left( \zerov, \Hm' \Sigmam_{X\!X} \Hm'^{\sf T}\right).
\end{IEEEeqnarray}

In the following, we first characterize how the incomplete admittance information impacts the attack covariance matrix 
$\Hm' \Sigmam_{X\!X} \Hm'^{\sf T}$. 
Using the obtained results, we then analyze the resulting attack performance.

\subsection{Incompleteness in Attack Covariance Matrix}
Firstly, we provide an alternative expression for the linearized observation model in \eqref{Equ:DCSE}, 
which follows directly by combining \eqref{Equ:DCSE} with \eqref{Equ:H}.

\begin{lemma}\label{Lemma:EquModel}

The linearized observation model in \eqref{Equ:DCSE} is equivalent to the composition of 
\begin{IEEEeqnarray}{rl}
\bar{X}^l &= \Am X^{n}   \qquad \qquad \qquad \textnormal{and} \label{Equ:LemmaEquModel1}\\
Y^{m} &= \Jm\Dm \bar{X}^l + Z^{m}.
\end{IEEEeqnarray}
\end{lemma}

Under the assumption that $X^{n} \sim \Nc \left(\zerov, \Sigmam_{X\!X} \right)$, we can obtain the following proposition from 
Lemma \ref{Lemma:EquModel}.
\begin{proposition}\label{Pro:Dis}
If the state variables $X^{n}$ follow a multivariate Gaussian distribution $\Nc \left(\zerov, \Sigmam_{X\!X} \right)$, it holds that 
\begin{IEEEeqnarray}{rl}
\bar{X}^l &\sim \Nc \left(\zerov, \Am\Sigmam_{X\!X} \Am^{\sf T}\right) \qquad \textnormal{and} \label{Equ:barXDis}\\
Y^{m} &\sim \Nc \left(\zerov, \Jm \Dm \Am\Sigmam_{X\!X} \Am^{\sf T} \Dm^{\sf T} \Jm^{\sf T} +\sigma^2 \Id \right) \label{Equ:YMDis}.  \IEEEeqnarraynumspace
\end{IEEEeqnarray}
\end{proposition}

The following lemma shows the impact of the incomplete admittance information on the covariance matrix of $\bar{A}^m$.
\begin{lemma}\label{Lemma:Mismatch}
The covariance matrix of $\bar{A}^m$ in \eqref{Equ:A_bar} is given by 
\begin{IEEEeqnarray}{rl}
\Hm' \Sigmam_{X\!X} \Hm'^{\sf T} &= \Hm \Sigmam_{X\!X} \Hm^{\sf T} + \Jm \Dm  \Deltam\Dm^{\sf T} \Jm^{\sf T}, \IEEEeqnarraynumspace \squeezeequ \label{Equ:LemmaMismatch1}\\
& = \Jm  \Dm \Am\Sigmam_{X\!X}  \Am^{\sf T}\Dm^{\sf T} \Jm^{\sf T}+ \Jm \Dm  \Deltam\Dm^{\sf T} \Jm^{\sf T} \label{Equ:Mismatch1}\\
& = \Jm  \Dm \left(\Am\Sigmam_{X\!X}  \Am^{\sf T} +\Deltam\right)\Dm^{\sf T} \Jm^{\sf T} \label{Equ:Mismatch2}
\end{IEEEeqnarray}
where $\Deltam$ is given by 
\begin{IEEEeqnarray}{c}\label{Equ:Delta} 
\Deltam  = \Phim \Am \Sigmam_{X\!X} \Am^{\sf T} +  \Am \Sigmam_{X\!X} \Am^{\sf T} \Phim^{\sf T } + \Phim \Am \Sigmam_{X\!X} \Am^{\sf T} \Phim^{\sf T } \IEEEeqnarraynumspace
\end{IEEEeqnarray}
with $\Phim$ in \eqref{Equ:alpha}.
\end{lemma}

\begin{proof}
The proof is presented in Appendix \ref{Sec:ProofMismatch}. 
\end{proof}

From Lemma \ref{Lemma:Mismatch}, we can also know that $\Am\Sigmam_{X\!X}  \Am^{\sf T} +\Deltam$ is always a PSD matrix. 
\begin{lemma}\label{Lemma:PSD}
$\Am\Sigmam_{X\!X}  \Am^{\sf T} +\Deltam$ is a PSD matrix. 
\end{lemma}
\begin{proof}
The proof is presented in Appendix \ref{Sec:ProofPSD}. 
\end{proof}

Combining the results in Lemma \ref{Lemma:EquModel}, Lemma \ref{Lemma:Mismatch} and Lemma \ref{Lemma:PSD} yields the following theorem.

\begin{theorem}\label{theorem:EquMismatch}
For information-theoretic attacks in Section \ref{Subsec:ITA}, the following cases are equivalent in terms of their effect on the attack covariance matrix. 
\begin{itemize}
\item Incomplete Admittance Information Case: the attacker does not know the branch admittance matrix perfectly, i.e. the attacker assumes it to be $\Dm'$ in \eqref{Equ:DBAMTD} instead of the true $\Dm$ in \eqref{Equ:D}, which introduces an additional term $\Hm' \Sigmam_{X\!X} \Hm'^{\sf T} - \Hm \Sigmam_{X\!X} \Hm^{\sf T} $ in the covariance matrix of the attack vector. 
\item Incomplete Covariance Matrix Case: 
the attacker does not know the covariance matrix of $\bar{X}^l$ in \eqref{Equ:LemmaEquModel1} perfectly, i.e. the attacker assumes it to be $\Am\Sigmam_{X\!X}\Am^{\sf T} + \Deltam$ instead of the true $\Am\Sigmam_{X\!X}\Am^{\sf T}$, which introduces an additional term $\Jm \Dm \Deltam \Dm^{\sf T} \Jm^{\sf T}$ in the covariance matrix of the attack vector.
\end{itemize}
\end{theorem}

For the covariance matrix of information-theoretic attacks, Theorem \ref{theorem:EquMismatch} shows that the imperfect attack covariance matrix induced by incomplete admittance information can be considered as that induced by incomplete information about covariance matrix of $\bar{X}^l$, in which there is a one-to-one mapping.
Thus, the effect of incomplete admittance information can be recast as a perturbation in the covariance matrix of $\bar{X}^l$, leading to an identical deviation in the attack covariance matrix.

While it would be ideal to interpret the incompleteness of the admittance matrix as an equivalent incompleteness in the covariance matrix of the state variables $X^n$, this equivalence does not generally hold. 
Specifically, expressing the discrepancy $\Deltam $ as a quadratic form
\begin{IEEEeqnarray}{c}
\Deltam = \Am \Xm \Am^{\sf T} \label{Equ:Ideal_Mismatch}
\end{IEEEeqnarray}
does not guarantee the existence of a valid matrix 
$\Xm$ that satisfies the equation for a given $\Deltam$ \cite[Section 13.2.1]{seber_matrix_2008}.
Therefore, it is in general impossible to interpret the effect of admittance information incompleteness  as an equivalent incompleteness in the covariance matrix of the state variables $X^n$.

\subsection{Attack Performance Evaluation}

When the attacker constructs the information-theoretic attacks as \eqref{Equ:A_bar}, the resulting measurements under attack are given by 
\begin{IEEEeqnarray}{c}\label{Equ:YBarA}
\bar{Y}_{A}^m = \Hm X^{n} + Z^{m} + \bar{A}^m, 
\end{IEEEeqnarray}
whose distribution is characterized in the following lemma. 

\begin{lemma}\label{Lemma:YBarADis}
The distribution of $\bar{Y}_{A}^m$ is given by 
\begin{IEEEeqnarray}{c}\label{Equ:YBarADis}
\bar{Y}_{A}^m \sim \Nc \left( \zerov, \Sigmam_{\bar{Y}_A\bar{Y}_A}\right),
\end{IEEEeqnarray}
where
\begin{IEEEeqnarray}{rl}
\Sigmam_{\bar{Y}_A\bar{Y}_A} &= \Sigmam_{YY} + \Hm' \Sigmam_{X\!X}\Hm'^{\sf T} \\
& = \Jm \Dm \left(2\Am\Sigmam_{X\!X} \Am^{\sf T} + \Deltam \right)\Dm^{\sf T} \Jm^{\sf T} +\sigma^2\Id
\end{IEEEeqnarray}
with $\Deltam$ in \eqref{Equ:Delta}. 
\end{lemma}
\begin{proof}
The proof is presented in Appendix \ref{Sec:ProofYBarADis}.
\end{proof}

Recall from Section \ref{Subsec:ITA} that the stealthiness and destructiveness of information-theoretic attacks are characterized by 
\begin{IEEEeqnarray}{c}\label{Equ:KLPD}
D\left(P_{\bar{Y}_{A}^m} \| P_{Y^m}\right) \quad \textnormal{and} \quad 
I\left(X^{n};\bar{Y}^{m}_{A}\right), \IEEEeqnarraynumspace
\end{IEEEeqnarray}
respectively, where the distributions of $\bar{Y}_{A}^m$ and $Y^m$ are provided in Lemma \ref{Lemma:YBarADis} and \eqref{Equ:YMDis}, respectively.
Substituting the expression of the KL divergence and the mutual information between multivariate Gaussian distributions \cite[Proposition 1\&2]{Sun_Stealth_2020} yields 
\begin{IEEEeqnarray}{rl}
D\left(P_{\bar{Y}_{A}^m} \| P_{Y^m}\right) & = \frac{1}{2}\left(-\log\left| \Id +\Sigmam_{YY}^{-1} \Tm\right| + \trace\left( \Sigmam_{YY}^{-1} \Tm\right)\right) \squeezeequ \IEEEeqnarraynumspace \\
& = \frac{1}{2}\!\left(-\log \left| \Id +\Sm^{\frac{1}{2}} \Tm\Sm^{\frac{1}{2}}\right| + \trace\left( \Sm^{\frac{1}{2}} \Tm\Sm^{\frac{1}{2}}\right)\!\right) \label{Equ:KL} \squeezeequ \IEEEeqnarraynumspace
\end{IEEEeqnarray}
and 
\begin{IEEEeqnarray}{rl}
I\left(X^{n};\bar{Y}^{m}_{A}\right) &= \frac{1}{2}\log\left| \Id + \left( \sigma^2\Id + \Tm\right)^{-1}\Hm\Sigmam_{X\!X}\Hm^{\sf T}\right| \IEEEeqnarraynumspace\\
& = \frac{1}{2}\log \left| \Id + \Um^{\frac{1}{2}}\left( \sigma^2\Id + \Tm\right)^{-1}\Um^{\frac{1}{2}}\right|, \label{Equ:DeltaCon1}
\end{IEEEeqnarray}
where 
\begin{IEEEeqnarray}{c}
\Tm \eqdef  \Jm  \Dm \left(\Am\Sigmam_{X\!X}  \Am^{\sf T} +\Deltam\right)\Dm^{\sf T} \Jm^{\sf T},\label{Equ:reuse1}
\end{IEEEeqnarray}
$\Um \eqdef \Hm\Sigmam_{X\!X}\Hm^{\sf T}$, and $\Sm \eqdef \Sigmam_{YY}^{-1}$. 
Note that from Lemma \ref{Lemma:Mismatch}, it is known that $\Tm$ is the covariance matrix of attacks, i.e. of $\bar{A}^{m}$, which is determined by the incomplete admittance information; and $\sigma^2$, $\Tm$, and $\Sm$ are constant or constant matrices. 

Note that $\Sm$ and $\Um$ are symmetric matrices. 
Hence, it can be known from the following lemma that $\Sm^{\frac{1}{2}} \Tm\Sm^{\frac{1}{2}}$ in \eqref{Equ:KL} and $\Um^{\frac{1}{2}}\left( \sigma^2\Id + \Tm\right)^{-1}\Um^{\frac{1}{2}}$ in \eqref{Equ:DeltaCon1} are PSD matrices.

\begin{lemma}\label{Lemma:PSD-G}
Let $\Mm  \in \Rc^{m \times m}$ and $\Xm \in \Sc^{+}_{m}$. 
Then it holds that $\Mm \Xm \Mm^{\sf T}$ is a PSD matrix. 
\end{lemma}
\begin{proof}
The proof is presented in Appendix \ref{Sec:ProofPSD-G}. 
\end{proof}

\section{Incomplete Admittance Information Induced Performance Degradation}\label{Sec:IAIIPD}

\subsection{Theoretical Analysis}\label{Sec:TA1}

In this section, we characterize the performance degradation induced by incomplete admittance information. 
Specifically, the attack performance of $\bar{A}^{m}$ in \eqref{Equ:A_bar} is compared with that of the optimal attack $\tilde{A}^{m}$ in \eqref{Equ:A_tilde} to show the performance degradation. 
In particular, we focus on two distinct operational regimes: (i) stealthiness intensifies while destructive potential diminishes and (ii) destructiveness increases while stealth capability weakens.

Prior to presenting the main result, Lemma \ref{Lemma:logdet_mono} and Lemma \ref{Lemma:CND} are introduced to establish the matrix-monotonicity and/or convexity properties of the stealthiness and destructiveness objective in \eqref{Equ:KL} and \eqref{Equ:DeltaCon1}. 
It is worth mentioning that the matrix-monotonicity is defined with respect to the Loewner order, which is a generalized inequality \cite[Example 2.15]{boyd_convex_2004}.

\begin{lemma}\label{Lemma:logdet_mono}
Let $\Mm \in \Rc^{m \times m}$.
Function $\log \left| \Id + \Mm\Xm^{-1}\Mm^{\sf T}\right|$ is a matrix-nonincreasing function of $\Xm$ on $\Sc^{m}_{++}$. 
\end{lemma}

\begin{proof}
The proof is presented in Appendix \ref{Sec:ProoflogdetMono}. 
\end{proof}

\begin{lemma}\label{Lemma:CND}
Function $-\log |\Id + \Xm| + \trace(\Xm)$ is a convex and matrix-nondecreasing function of $\Xm$ on $\Sc_{+}^{m}$.  
Furthermore, it holds that 
\begin{IEEEeqnarray}{c}
\zerov = \underset{\mathbf{X}\in \mathcal{S}_{+}^{m}}{\textnormal{arg} \min} -\log |\Id + \Xm| + \trace(\Xm). 
\end{IEEEeqnarray}

\end{lemma}
\begin{proof}
The proof is presented in Appendix \ref{Sec:ProofCND}.
\end{proof}

Using the results in Lemma \ref{Lemma:logdet_mono} and Lemma \ref{Lemma:CND}, we can obtain the following theorem. 

\begin{theorem}\label{Lemma:Delta_Con}
Let  $\tilde{Y}_A^m \eqdef Y^{m} + \tilde{A}^m$ with $\tilde{A}^m$ in \eqref{Equ:A_tilde}. 
If it holds for $\Deltam$ in \eqref{Equ:Delta} that 
\begin{IEEEeqnarray}{c}\label{Equ:Sun1}
\Deltam \succeq \zerov, 
\end{IEEEeqnarray}
it holds that 
\begin{IEEEeqnarray}{rl}
D\left(P_{\bar{Y}_{A}^m} \| P_{Y^m}\right) & \geq D\left(P_{\tilde{Y}_A^m} \| P_{Y^m}\right)  \quad \textnormal{and}\\
I\left(X^{n};\bar{Y}^{m}_{A}\right) &\leq I\left(X^{n};\tilde{Y}^{m}_{A}\right).
\end{IEEEeqnarray}
If it holds for $\Deltam$ in \eqref{Equ:Delta} that 
\begin{IEEEeqnarray}{c}\label{Equ:Sun2}
\Deltam \preceq \zerov, 
\end{IEEEeqnarray}
it holds that 
\begin{IEEEeqnarray}{rl}
D\left(P_{\bar{Y}_{A}^m} \| P_{Y^m}\right) &\leq D\left(P_{\tilde{Y}_A^m} \| P_{Y^m}\right) \quad \textnormal{and} \\
I\left(X^{n};\bar{Y}^{m}_{A}\right) &\geq I\left(X^{n};\tilde{Y}^{m}_{A}\right).
\end{IEEEeqnarray}
\end{theorem}

\begin{proof}
The proof is presented in Appendix \ref{Sec:ProofDeltaCon}.
\end{proof}

Note that $\tilde{Y}_A^m$ denotes the measurements under attack when the attacker has complete knowledge of the admittance matrix, i.e. the optimal attack scenario described in \eqref{Equ:A_tilde}. 
To that end, the quantities $D\left(P_{\tilde{Y}_A^m} \| P_{Y^m}\right) $ and $I\left(X^{n};\tilde{Y}^{m}_{A}\right)$ represent the optimal tradeoff between stealthiness and destructiveness. 
Hence, Theorem \ref{Lemma:Delta_Con} shows that if $\Deltam \succeq \zerov$, both the probability of detection and the level of induced disruption increase (or at least do not decrease), which implies that destructiveness is enhanced at the cost of reduced stealth. 
Conversely, if $\Deltam \preceq \zerov$, both metrics decrease (or at least do not increase), which implies that stealthiness is improved while the attack’s disruptive potential diminishes.

The intuition behind the result in Theorem \ref{Lemma:Delta_Con} is straightforward. 
When $\Deltam \succeq \zerov$, it holds that $\Am \Sigmam_{X\!X} \Am^{\sf T} + \Deltam \succeq \Am \Sigmam_{X\!X} \Am^{\sf T}$. 
In this case, as indicated by \eqref{Equ:Mismatch2}, the incomplete admittance information makes the attacker increase the power of the injected attacks. 
As a result, the probability of detection and the level of induced disruption increase. 
Similarly, when $\Deltam \preceq \zerov$, it holds that $\Am \Sigmam_{X\!X} \Am^{\sf T} + \Deltam \preceq \Am \Sigmam_{X\!X} \Am^{\sf T}$, which implies that the attacker reduces the power of the injected attacks. 
Consequently, the probability of detection and the level of induced disruption decrease. 
Overall, Theorem \ref{Lemma:Delta_Con} reveals that the ``sign'' of $\Deltam$ ($\Deltam \succeq \zerov$ or $\preceq \zerov$)  directly determines whether incomplete admittance information amplifies or suppresses attack power, thereby jointly affecting attack stealthiness and destructiveness in a consistent manner.
To that end, $\Deltam$ can be viewed by the operator as a diagnostic indicator of how admittance uncertainty biases the attacker’s effective action.

It is worth mentioning that the proof of Theorem \ref{Lemma:Delta_Con} does not rely on any specific structure of $\Deltam$;  in other words, $\Deltam$ is not required to conform to the form given in \eqref{Equ:Delta}.
In fact, the result in Theorem \ref{Lemma:Delta_Con} naturally extends to any matrix $\bar{\Deltam} \succeq -\Am \Sigmam_{X\!X}\Am^{\sf T}$, which ensures that $\Am\Sigmam_{X\!X}  \Am^{\sf T} +\bar{\Deltam}$ is a PSD matrix (see Lemma \ref{Lemma:PSD}). 
Consequently, if $\bar{\Deltam} \succeq \zerov$,  both the probability of detection and the level of disruption increase (or at least do not decrease); 
and if $\bar{\Deltam} \preceq \zerov$, both metrics decrease (or at least do not increase). 
However, under this generalized setting, there is no direct way to interpret the introduction of $\bar{\Deltam}$ as arising from incomplete admittance information. 
One possible interpretation is that $\bar{\Deltam}$ is induced by changes in system topology, such as modifications of the matrix 
$\Am$ in \eqref{Equ:BBIM} due to transmission line switching \cite{morrow_2012_topology, wang_2015_effects}. 
Unfortunately, this interpretation is speculative and has not been formally established.

Theorem \ref{Lemma:Delta_Con} considers the condition for $\Deltam$ in \eqref{Equ:Delta}, and $\Deltam$ is determined by $\Phim$ in \eqref{Equ:alpha}.
The following lemma establishes a sufficient condition for $\Phim$ such that $\Deltam \succeq \zerov$ or $\Deltam \preceq \zerov$. 

\begin{lemma}\label{Lemma:SC_Delta}
Let $\phiv = \left(\phi_1, \ldots, \phi_k, 0, \ldots, 0\right)^{\sf T} \in \Rc^{l}$ with $\phi_i$ in \eqref{Equ:DBAMTD1}, and $\mathbf{1} = \{1, \ldots, 1 \}^{\sf T}$ be a vector of proper dimension whose elements all equal to $1$. If 
\begin{IEEEeqnarray}{c}
\phiv^{\sf T} \mathbf{1} - \sqrt{\|\phiv\|_2^{2} \| \mathbf{1}\|_2^{2}} \geq 0, \label{Equ:Mon1}
\end{IEEEeqnarray}
it holds that $\Deltam \succeq \zerov$; if 
\begin{IEEEeqnarray}{c}
\phiv^{\sf T} \phiv + \phiv^{\sf T} \mathbf{1} + \sqrt{\|\phiv\|_2^{2} \|\mathbf{1}\|_2^{2}} \leq 0,  \label{Equ:Mon2}
\end{IEEEeqnarray}
it holds that $\Deltam \preceq \zerov$. 

\end{lemma}

\begin{proof}
The proof is presented in Appendix \ref{Sec:ProofDeltaSC}. 
\end{proof}

Lemma \ref{Lemma:SC_Delta} establishes that if the condition in \eqref{Equ:Mon1} holds, $\Deltam \succeq \zerov$, then from Theorem \ref{Lemma:Delta_Con}, the probability of detection and the disruption are increased (or at least do not decrease). 
Similarly, if the condition in \eqref{Equ:Mon2} holds, $\Deltam \preceq \zerov$, and from Theorem \ref{Lemma:Delta_Con}, the probability of detection and the disruption are decreased (or at least do not decrease). 
It is worth mentioning that the condition in Lemma \ref{Lemma:SC_Delta} serves as sufficient condition for $\Deltam \succeq \zerov$ or $\Deltam \preceq \zerov$.
Therefore, the result in Theorem \ref{Lemma:Delta_Con} is more general than the result in Lemma \ref{Lemma:SC_Delta}. 

However, the conditions in \eqref{Equ:Mon1} and \eqref{Equ:Mon2} are not easy to achieve. 
Specifically, it follows from the Cauchy–Schwarz inequality \cite[12.1(a)]{seber_matrix_2008} that 
\begin{IEEEeqnarray}{c}
\phiv^{\sf T} \mathbf{1} / \sqrt{\|\phiv\|_2^{2}\| \mathbf{1}\|_2^{2}} \in [-1,1],
\end{IEEEeqnarray}
which implies that $\phiv^{\sf T} \mathbf{1} - \sqrt{\|\phiv\|_2^{2} \| \mathbf{1}\|_2^{2}} \leq 0$ and $\phiv^{\sf T} \phiv + \phiv^{\sf T} \mathbf{1} + \sqrt{\|\phiv\|_2^{2}\| \mathbf{1}\|_2^{2}} \geq 0$. 
Therefore, a necessary condition for satisfying \eqref{Equ:Mon1} and \eqref{Equ:Mon2} is that $\phiv$ and $\onev$ are linearly dependent. 
This indicates that the admittances of all the branches must be perturbed by the same ratio in order to fulfill the conditions in \eqref{Equ:Mon1} and \eqref{Equ:Mon2}.

The following proposition can be obtained by selecting a vector $\phiv$ that is linearly dependent on $\onev$, or equivalently, setting $\Phim = \beta \Id$ for some $\beta \in \Rc$. 

\begin{proposition}\label{Pro:2}
Assume that it holds for $\Phim$ in \eqref{Equ:alpha} that $\Phim = \beta \Id$ with $\beta \in \Rc$. 
If $\beta \geq 0 \ \textnormal{or} \ \beta \leq -2$, it holds that $\Deltam \succeq \zerov$; if $ -2 \leq \beta \leq 0$, it holds that $\Deltam \preceq \zerov$.
\end{proposition}

\subsection{Maximal Degradation in Probability of Detection}\label{Sec:MD1}

For scenarios in which the sufficient condition in \eqref{Equ:Sun1} or \eqref{Equ:Sun2} is not satisfied,  we shift our focus to maximizing the performance degradation of the attack. 
Specifically, we aim to maximize the probability of detection. 
The rationality follows from the fact that: if the operator detects anomalies in the system, they can take appropriate actions to mitigate the distortions caused by the attack. 
Conversely, if the system appears normal, the operator is unlikely to intervene, allowing the attack to persist unchallenged.

Note that the admittance information available to the attacker cannot deviate arbitrarily from the true values, as physical constraints prevent the branch admittances from becoming unboundedly large or small. 
Let $\phi_{i,max}$ and $\phi_{i,min}$ denote the upper and lower bounds on the incompleteness of branch admittance $b_i$, respectively. 
Specifically, it holds for the incompleteness parameter $\phi_i$ in \eqref{Equ:DBAMTD1} that
\begin{IEEEeqnarray}{c}\label{Equ:UBLW1}
\phi_{i, min} \leq \phi_i \leq \phi_{i, max}, \ \forall i \in \{1, \ldots,k \}. 
\end{IEEEeqnarray}

Under such a setting, maximizing the degradation in probability of detection can be formulated as the following optimization problem. 
\begin{IEEEeqnarray}{rl}
\textnormal{\bf P1}: \ \ \underset{ {\bf \Phi} \in \Dc^{l}}{\textnormal{max}} & \ -\log \left| \Id + \Sm^{\frac{1}{2}} \Tm \Sm^{\frac{1}{2}} \right| + \trace \left( \Sm^{\frac{1}{2}} \Tm \Sm^{\frac{1}{2}}\right) \IEEEeqnarraynumspace \label{Equ:obj}\\
\textnormal{s.t.} & \ \textnormal{Constraint in \eqref{Equ:Delta}}, \ \\
& \ \Tm =  \Jm  \Dm \left(\Am\Sigmam_{X\!X}  \Am^{\sf T} +\Deltam\right)\Dm^{\sf T} \Jm^{\sf T} , \label{Equ:Tm} \IEEEeqnarraynumspace \\
&\  \Phim^{min} \preceq \Phim \preceq  \Phim^{max}, \label{Equ:reuse4}
\end{IEEEeqnarray}
where objective function \eqref{Equ:obj} follows from \eqref{Equ:KL}; the constraint \eqref{Equ:Tm} follows from \eqref{Equ:reuse1}; and \eqref{Equ:reuse4} follows from \eqref{Equ:UBLW1}, in which  
\begin{IEEEeqnarray}{rl}
\Phim^{min} &= \textnormal{diag} \left(\phi_{1,min}, \ldots, \phi_{k,min}, 0, \ldots, 0\right) \quad \textnormal{and} \IEEEeqnarraynumspace \label{Equ:reuse2}\\
\Phim^{max} &= \textnormal{diag} \left(\phi_{1,max}, \ldots, \phi_{k,max}, 0, \ldots, 0\right). \label{Equ:reuse3}
\end{IEEEeqnarray}
Note that the constraint \eqref{Equ:reuse4} is equivalent to $ \Phim^{min} \leq \Phim \leq  \Phim^{max}$ under the setting in \eqref{Equ:alpha} and \eqref{Equ:UBLW1}.

The following theorem shows the convexity of problem $\bf P1$ and a reduce-sized set for the solution. 

\begin{theorem}\label{theorem:PDMax}
In $\bf P1$, a convex function of $\Phim$ is maximized within a convex set. 
Furthermore, the optimal solution to $\bf P1$, i.e. $\Phim^{\star}$, belongs to the set 
\begin{IEEEeqnarray}{c}\label{Equ:PDmax1}
\big\{\Phim \in \Dc^{l} : \phi_{i} \in \{\phi_{i,min}, \phi_{i,max} \}, \ \forall i \in \{1, \ldots,k\} \big\}, \IEEEeqnarraynumspace
\end{IEEEeqnarray}
where $k$ is defined in \eqref{Equ:alpha}.
\end{theorem}

\begin{proof}
The proof is presented in Appendix \ref{Sec:ProofPDMax}.
\end{proof}

\begin{algorithm}[t!]
\caption{Stealthiness Degradation Maximization }\label{Algo:greedy_1}
\begin{algorithmic}[1]
\Require system information matrices $\Jm$, $\Dm$, and $\Am$; dimension $l$; statistical matrices $\Sigmam_{X\!X}$ and $\Sm$; upper bound $\Phim^{max}$ and lower bound $\Phim^{min}$
\Ensure $\Phim$
\State Set $i= 1$ and $\Phim = \zerov$
\While{$ i \leq k$}
\If{$\phi_{i, min} = \phi_{i, max}$} 
    \State $\phi_{ii} = \phi_{i, min}$
\Else
    \State 
    Set $\Phim_1 = \Phim + \phi_{i, min}\Em_{ii}$, $\Phim_2 = \Phim + \phi_{i, max}\Em_{ii}$
    \State  Set  $\Deltam_1 = \Deltam(\Phim_1)$,  $\Deltam_2 =  \Deltam(\Phim_2)$
    \State Set  $\Tm_1 = \Tm(\Deltam_1)$,  $\Tm_2 = \Tm(\Deltam_2)$
    \State Set $D_1 = D(\Tm_1)$, $D_2 = D(\Tm_2)$
        \If{$D_1 \geq D_2$}
        \State $\phi_{ii} = \phi_{i, min}$
    \Else
        \State $\phi_{ii} = \phi_{i, max}$
    \EndIf
\EndIf 
\State Set $\Phim = \Phim + \phi_{ii}\Em_{ii}$
\State $i \gets i+1$
\EndWhile\label{euclidendwhile}
\end{algorithmic}
\end{algorithm}

Theorem \ref{theorem:PDMax} shows that the optimal solution to problem {\bf P1} lies at the vertices of the polyhedron defined by the constraint $\Phim^{min} \leq \Phim \leq  \Phim^{max}$.
The intuition behind this result is quite straightforward: 
in order to maximize the probability of detection caused by incomplete admittance information, the degree of incompleteness (i.e., deviation in the available admittance information) should be as large as permitted within the given bounds.

Although the feasible region is reduced without excluding the optimal solution in Theorem \ref{theorem:PDMax}, the cardinality of the set in \eqref{Equ:PDmax1} is still $2^k$, which implies that an exhaustive search for the optimal solution is computationally infeasible for large $k$. 
Moreover, maximizing a convex function within a convex set is generally an NP-hard problem \cite{ben_2022_algorithm}. 
To address this challenge, we propose the greedy algorithm in Algorithm \ref{Algo:greedy_1}, in which $\Em_{ii}$ is a matrix with a one in the $(i,i)$-th position and zeros elsewhere. 
And with a slight abuse of notation, the matrix $\Deltam$ in \eqref{Equ:Delta} and matrix $\Tm$ in \eqref{Equ:reuse1} are henceforth regarded as functions of matrices $\Phim$ and $\Deltam$, respectively; moreover, the KL divergence in \eqref{Equ:KL} is viewed as a scalar-valued function $D: \Rc^{m \times m} \to \Rc^{+}$.

The intuition behind Algorithm \ref{Algo:greedy_1} is as follows. 
According to Theorem \ref{theorem:PDMax}, the diagonal elements of $\Phim^{\star}$ are either $\phi_{i,min}$ or $\phi_{i,max}$ for all $i \in \{ 1, \ldots, k\}$. 
Therefore, in the $i$-th iteration of the algorithm, the value of $\phi_{ii}$ is selected based on which candidate ($\phi_{i,\min}$ or $\phi_{i,\max}$) yields a higher objective value.
Note that the solution obtained by Algorithm \ref{Algo:greedy_1} belongs to the set in \eqref{Equ:PDmax1}, but there is no theoretical guarantee that the solution is globally optimal.

\subsection{Practical Implementation}\label{Sec:PI}

In practical scenarios, the incomplete admittance information $\Phim$ in \eqref{Equ:alpha} can arise from various mechanisms, such as MTD. 
Specifically, the system operator intentionally alters the admittances of selected branches \cite{rahman_2014_moving} using flexible AC transmission system (FACTS) devices  \cite{hingorani_2007_facts}  to introduce incompleteness to admittance information.
In this setting, both the attacker and the operator initially have access to the Jacobian matrix $\Hm'$ in \eqref{Equ:NH} (or equivalently, the $\Dm'$ in \eqref{Equ:DBAMTD}) before the implementation of MTD. 
After MTD is applied, the admittance matrix is updated to $\Dm$ as defined in \eqref{Equ:D}, but the attacker is unaware of this and continues to believe that $\Dm'$ is the true admittance matrix. 

Note that from \eqref{Equ:DBAMTD1}, it holds for $i \in \{ 1, \ldots, k\}$ that 
\begin{IEEEeqnarray}{c}\label{Equ:PI1}
b_{i} = \left\{ 
\begin{tabular}{cl}
$\frac{1}{1+\phi_i} b_i'$, & \textnormal{if $\phi_{i} \neq -1$} ,\\
$0 $, & \textnormal{if $\phi_{i} = -1$}, \\
\end{tabular}
\right. 
\end{IEEEeqnarray}
in which $b_i'$ is the initial admittance of branch $i$ known by the attacker, or equivalently, the admittance of branch $i$ prior to the implementation of MTD; and $b_{i}$ is the admittance of branch $i$ after the implementation of MTD. 
To that end, for a given $\Phim$ with diagonal element $\phi_i$, the operator can adjust the admittance of the $i$-th branch to $b_{i}$ via \eqref{Equ:PI1} to achieve different MTD strategies.

\section{Further Extension}\label{Sec:FE}

In Section \ref{Sec:TA1}, we employed monotonicity arguments with respect to the Loewner order, which require that the matrix $\Deltam \succeq \zerov$ or  $\Deltam \preceq \zerov$, i.e. $\Deltam$ is PSD or negative semidefinite. 
In this section, we explicitly consider the case in which $\Deltam$ is indefinite and characterize the corresponding performance.
\subsection{Example}
We first establish a sufficient condition under which $\Deltam$ is indefinite, thereby providing further insight into the source of its indefiniteness. 

\begin{lemma}\label{Lemma:Ind}
Let $\Nm \eqdef \Am\Sigmam_{X\!X}  \Am^{\sf T}$, and $d_i \eqdef(1+\phi_i)^2 - 1$ with $\phi_i$ in  \eqref{Equ:alpha} for all $i\in \{ 1, \ldots, k\}$.
If there exist two indices $i$ and $j$ with $i\neq j$ such that
\begin{IEEEeqnarray}{rl}
d_i N_{ii} < 0   \ \ &\textnormal{and}   \ \ d_j N_{jj} > 0, \quad \textnormal{or} \label{Equ:Ind_con1} \\ 
d_i N_{ii} > 0    \ \ &\textnormal{and}  \ \  d_j N_{jj} < 0,\label{Equ:Ind_con2}
\end{IEEEeqnarray}
it holds that $\Deltam$ is an indefinite matrix. 
\end{lemma}
\begin{proof}
The proof is presented in Appendix \ref{Sec:ProofInd}.
\end{proof}

By Proposition \ref{Pro:Dis}, matrix $\Am\Sigmam_{X\!X}  \Am^{\sf T}$ (or equivalently, $\Nm$) is the covariance matrix of $\bar{X}^l$. 
Hence, its diagonal entries satisfy $N_{ii}\geq 0$ for all relevant indices $i$. 
Lemma \ref{Lemma:Ind} shows that if two diagonal contributions $d_iN_{ii}$ and $d_jN_{jj}$ have 
opposite signs, then $\Deltam$ is an indefinite matrix.

This condition admits a simple interpretation. 
Suppose that all diagonal entries of $\Am\Sigmam_{X\!X}\Am^{\sf T}$ are nonzero, i.e. all components 
of $\bar{X}^l$ have nonzero variance.
Since $d_i=(1+\phi_i)^2-1$, we have $d_i>0$ when $\phi_i>0$, and $d_i<0$ when $-2<\phi_i<0$.
Note that from \eqref{Equ:PI1}, it holds that $b_i < b_i'$ when $\phi_i>0$ and $b_i > b_i'$ when $-2<\phi_i<0$. 
Therefore, {\it if the operator decreases the admittance of one branch (i.e. by setting $\phi_i>0$) and moderately increases  another branch admittance (i.e. by setting $-2<\phi_j<0$)}, then the corresponding $d_iN_{ii}$ and $d_jN_{jj}$ have opposite signs. 
In this case, Lemma~\ref{Lemma:Ind} implies that {\it $\Deltam$ is indefinite}.


\subsection{Performance Degradation under Indefinite Case}

We next provide a method to characterize the performance when $\Deltam$ is indefinite.
Although no universal Loewner-order conclusion can be drawn when $\Deltam$ is indefinite, the effect of a given indefinite perturbation can still be diagnosed exactly by comparing the two performance metrics with
their complete-information benchmarks.

Define the two signed diagnostic indices
\begin{IEEEeqnarray}{c}
\eta_{\rm det}(\Deltam)\triangleq
D\!\left(P_{\bar{Y}_A^m}\|P_{Y^m}\right)-D\!\left(P_{\tilde{Y}_A^m}\|P_{Y^m}\right),
\end{IEEEeqnarray}
and
\begin{IEEEeqnarray}{c}
\eta_{\rm des}(\Deltam)\triangleq
I\!\left(X^n;\tilde{Y}_A^m\right)-I\!\left(X^n;\bar{Y}_A^m\right).
\end{IEEEeqnarray}
The first quantity measures the change in detectability (i.e. probability of detection); and the second measures the change in destructiveness, since a smaller mutual information corresponds to a stronger disruption of state estimation.

Using \eqref{Equ:KL} and \eqref{Equ:DeltaCon1},  these two quantities can be computed as
\begin{IEEEeqnarray}{c}
\eta_{\rm det}(\Deltam)=\frac{1}{2}\left[-\log\frac{\left|\Id+\Sm^{1/2}\Tm_{\Delta}\Sm^{1/2}\right|}{
\left| \Id+\Sm^{1/2}\Tm_{0}\Sm^{1/2}\right|}+
\operatorname{tr}\left(\Sm^{1/2}\Em_{\Delta}\Sm^{1/2}\right)\right], \supersqueezeequ \IEEEeqnarraynumspace
\end{IEEEeqnarray}
and
\begin{IEEEeqnarray}{c}
\eta_{\rm des}(\Deltam)=\frac{1}{2}\log\frac{\left|\Id+\Um^{1/2}(\sigma^2 \Id+\Tm_0)^{-1}\Um^{1/2}\right|}{\left|\Id+\Um^{1/2}(\sigma^2 \Id+\Tm_{\Delta})^{-1}\Um^{1/2}\right|},
\end{IEEEeqnarray}
where $\Tm_0 \triangleq \Um = \Jm\Dm\Am\Sigmam_{X\!X}\Am^{\sf T}\Dm^{\sf T}\Jm^{\sf T}$, $\Em_{\Delta} \triangleq \Jm\Dm\Deltam \Dm^{\sf T}\Jm^{\sf T}$, and $\Tm_{\Delta} \triangleq \Tm_0 + \Em_{\Delta}$.
Therefore, for an indefinite $\Deltam$, the signs of $\eta_{\rm det}(\Deltam)\) and \(\eta_{\rm des}(\Deltam)$ provide an exact diagnostic criterion: $\eta_{\rm det}(\Deltam)>0$ indicates increased detectability, i.e., reduced stealthiness; while $\eta_{\rm des}(\Deltam)>0$ indicates increased destructiveness. 

Thus, although Theorem \ref{Lemma:Delta_Con} provides universal monotonic guarantees only when
$\Deltam \succeq \zerov$ or $\Deltam \preceq \zerov$, the above indices provide an exact instance-wise diagnostic for determining whether detectability and destructiveness increase or decrease.

\section{Numerical Simulation}\label{Sec:Sim}

In this section, we present simulation results to evaluate the findings under practical state estimation settings. 
Specifically, we employ the IEEE test system framework provided by MATPOWER \cite{zimmerman_matpower:_2011}, and consider a DC state estimation scenario \cite{abur_power_2004, grainger_power_1994} to fulfill the linearized system in \eqref{Equ:DCSE} and the Jacobian matrix in \eqref{Equ:H}.
The simulation settings are the same as in \cite{Sun_information-theoretic_2017, Sun_Stealth_2020, sun_2019_learning, sun_2023_asymptotic}.
The covariance matrix of the state variables is assumed to be a Toeplitz matrix with exponential decay parameter $\rho$, which determines the correlation strength between different entries of the state variable vector.
And the Signal-to-Noise Ratio (SNR) of the power system  is defined as
\begin{equation}
\textnormal{SNR} \eqdef 10\log_{10}\left(\frac{\trace{(\Hm\Sigmam_{X\!X}\Hm^\textnormal{T}})}{m\sigma^2}\right).
\end{equation}

It is worth noting that the theoretical results presented in this paper hold for arbitrary state covariance matrices. 
However, demonstrating these results through simulations requires selecting a specific covariance model. While an empirical matrix would ideally capture the complex correlations induced by real-world network topology and power-flow constraints, constructing one is practically difficult.
Standard IEEE test cases lack historical time-series measurements, and public load datasets lack corresponding network information. 
To address this, we employ the Toeplitz covariance model as a controlled and reproducible benchmark, which allows us to systematically tune the correlation strength among state variables.
Although the exact numerical outcomes may vary depending on the specific choice of covariance model, SNR regime, or perturbation pattern, the underlying analytical conclusions and regime characterizations remain unchanged. 
 Ultimately, these numerical results are included to complement the theoretical analysis by examining representative scenarios.

\subsection{Simulation of Results in Section \ref{Sec:TA1}}\label{Sec:Sim1}
\begin{figure}[t!]
\centering
\includegraphics[scale=0.47]{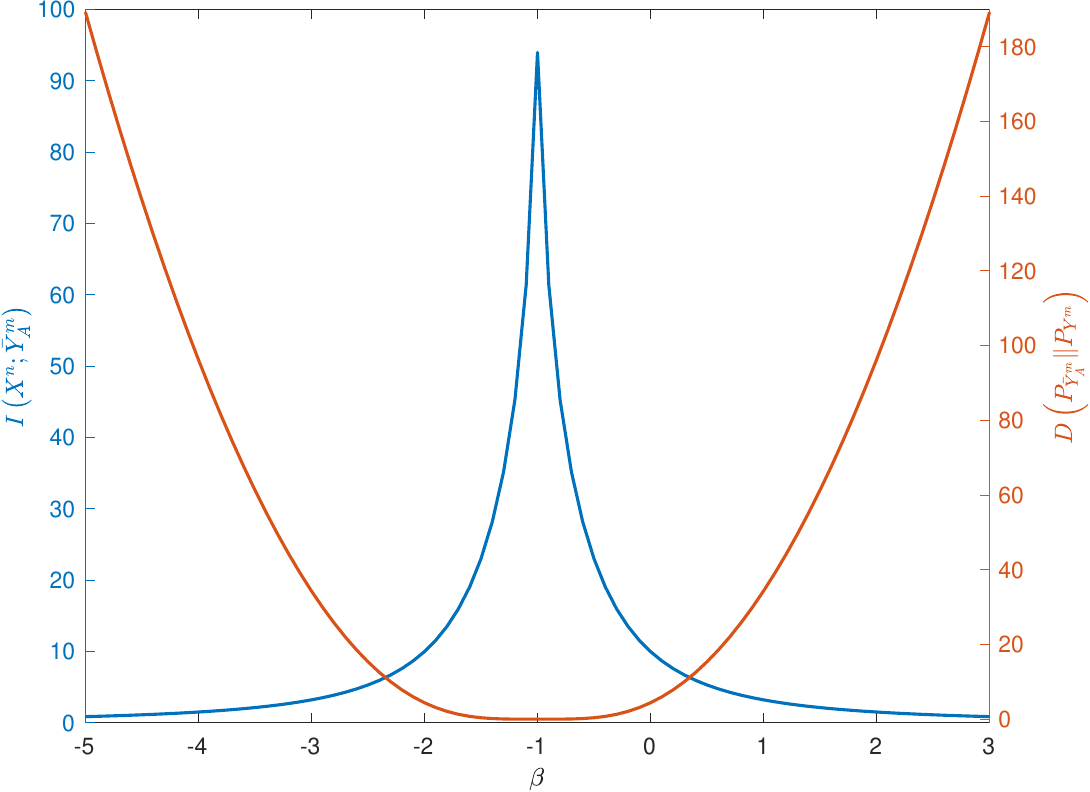}
\caption{Performance of the attack in \eqref{Equ:A_bar} on IEEE 30-Bus test system when $\textnormal{SNR}=30 \ \textnormal{dB}$ and $\rho=0.5$.}
\label{Fig:MIKL_30_SNR30_rho05}
\end{figure}

Fig. \ref{Fig:MIKL_30_SNR30_rho05} illustrates the performance of the attack in \eqref{Equ:A_bar} on IEEE 30-Bus test system when $\textnormal{SNR}=30\ \textnormal{dB}$ and $\rho=0.5$, in which $\Phim$ in \eqref{Equ:alpha} is given by $\Phim = \beta \Id$ (i.e. $\Phim$ satisfies the condition specified in Proposition \ref{Pro:2}).
From \eqref{Equ:Delta}, it is known that $\Deltam = \left(2\beta + \beta^2\right) \Am \Sigmam_{X\!X}\Am^{\sf T}$.
To that end, 
\begin{IEEEeqnarray}{c}\label{Equ:Example1}
\Deltam 
\left\{
\begin{array}{ll}
         \succeq \zerov, & \textnormal{if } \beta \geq 0 \textnormal{ or } \beta \leq -2;\\
        \preceq \zerov, &  \textnormal{if } -2 < \beta < 0.
\end{array}
\right. \label{Equ:example1}
\end{IEEEeqnarray}
Furthermore, when $\beta = -1$, it follows from Lemma \ref{Lemma:Mismatch} that the covariance matrix of $\bar{A}^{m}$ in \eqref{Equ:A_bar} is a matrix of all zero elements, which means that no attack is injected into the system. 
And when $\beta = -2$ or $0$, it follows from \eqref{Equ:Delta} that $\Deltam = \zerov$ and  the covariance matrix of $\bar{A}^{m}$ in \eqref{Equ:A_bar} achieves the optimal tradeoff between the attack stealthiness and
destructiveness.

\begin{figure}[t!]
\centering
\includegraphics[scale=0.47]{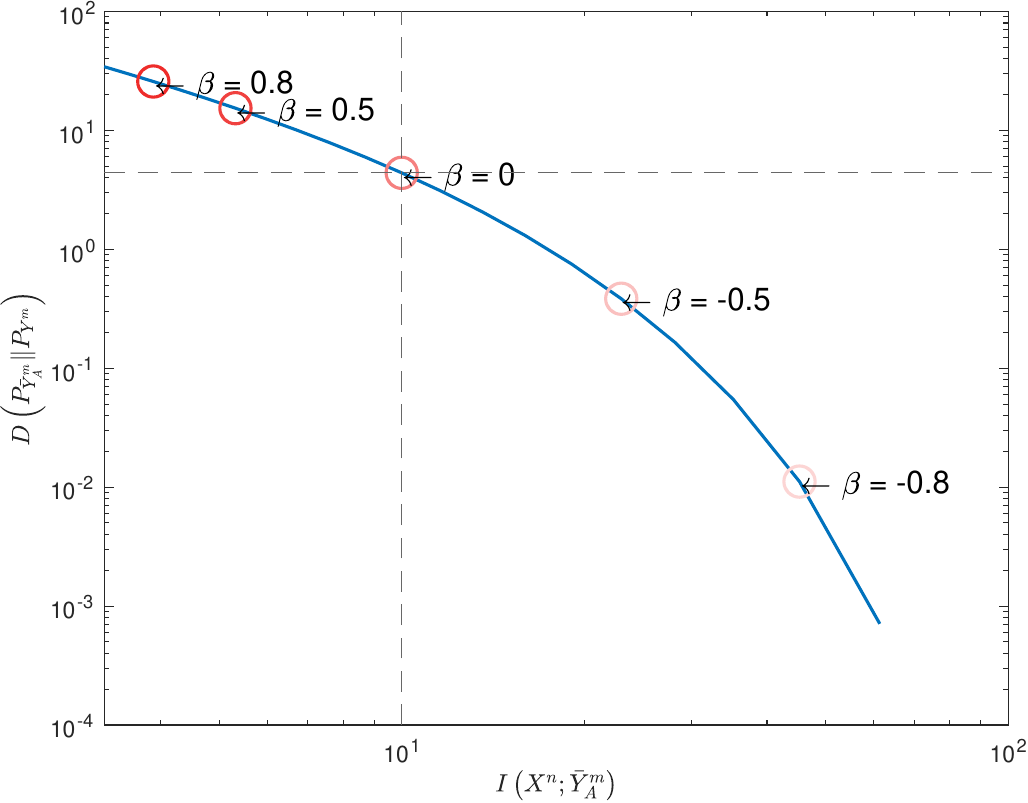}
\caption{Tradeoff change on IEEE 30-Bus test system when $\beta \in \left[-1, 1\right]$, $\textnormal{SNR}=30 \ \textnormal{dB}$, and $\rho=0.5$.}
\label{Fig:MIKLTradeOff_30_SNR30_rho05}
\end{figure}

To that end, when $\beta \in \left[-2,0\right]$, the probability of detection is reduced but amount of the information obtained by the operator increases, compared with the case with complete admittance information. 
Conversely, when $\beta \leq -2$ or $\beta \geq 0$, the operator gains less information, but the probability of detection increases.
These observations validate the result in Theorem \ref{Lemma:Delta_Con} and Lemma \ref{Lemma:SC_Delta}. 
Fig. \ref{Fig:MIKLTradeOff_30_SNR30_rho05} further shows the change in the tradeoff between the probability of detection and the information obtained by the operator when $\beta \in \left[-1, 1\right]$, in which $\beta = 0$  corresponds to the optimal tradeoff and darker red circles indicate higher values of $\beta$.

Additionally, it can be found from Fig. \ref{Fig:MIKL_30_SNR30_rho05} that both the mutual information in \eqref{Equ:DeltaCon1}  and the KL divergence in \eqref{Equ:KL} are symmetric with respect to $\beta = -1$; and 
the mutual information reaches the maximum value and the KL divergence in \eqref{Equ:KL} attains the minimum value at $\beta = -1$. 
As $\beta$ deviates from $-1$, the KL divergence increases and exhibits a convex relationship with $\beta$, which validates the result in Lemma \ref{Lemma:CND}.  
In contrast, although the mutual information decreases as $\beta$ moves away from $-1$, it does not follow a convex or concave pattern.
This observation implies that the function in Lemma \ref{Lemma:logdet_mono} is neither a convex nor concave function with respect to $\beta$.

It is worth mentioning that the conclusions and observations in this section hold consistently across various IEEE test systems, SNR levels, and correlation strength $\rho$. 
Therefore, additional simulation results under alternative settings are omitted for brevity.

\subsection{Simulation of Results in Section \ref{Sec:MD1}} 
Intuitively, the performance of Algorithm \ref{Algo:greedy_1} is determined by the number of branches with incomplete admittance information and the bounds (upper and lower bounds) on their incompleteness, specifically, by $|{\cal E_{I}}|$ with ${\cal E_{I}}$ in \eqref{Equ:DBAMTD} and the values of $\phi_{i, min}$ and $\phi_{i, max}$ in \eqref{Equ:UBLW1}.
To quantify the overall degree of incompleteness across branches, we use the Frobenius norm of the difference between $\Phim^{max}$ in \eqref{Equ:reuse2} and $\Phim^{min}$ in \eqref{Equ:reuse3} as an index. 
Specifically, we use $\alpha \in \Rc^{+}$ with
\begin{IEEEeqnarray}{c}
\alpha \eqdef \| \Phim^{max} - \Phim^{min} \|_F
\end{IEEEeqnarray}
to measure the overall incompleteness. 
Given the fact that $\Phim^{max}$ and $\Phim^{min}$ are both diagonal matrices with diagonal elements $\phi_{i, max}$ and $\phi_{i, min}$, it holds for $\alpha$ that 
\begin{IEEEeqnarray}{c}
\alpha = \sqrt{\sum_{i=1}^k \left(\phi_{i, max} - \phi_{i, min}\right)^2}.
\end{IEEEeqnarray}

\begin{figure}[t!]
\centering
\includegraphics[scale=0.48]{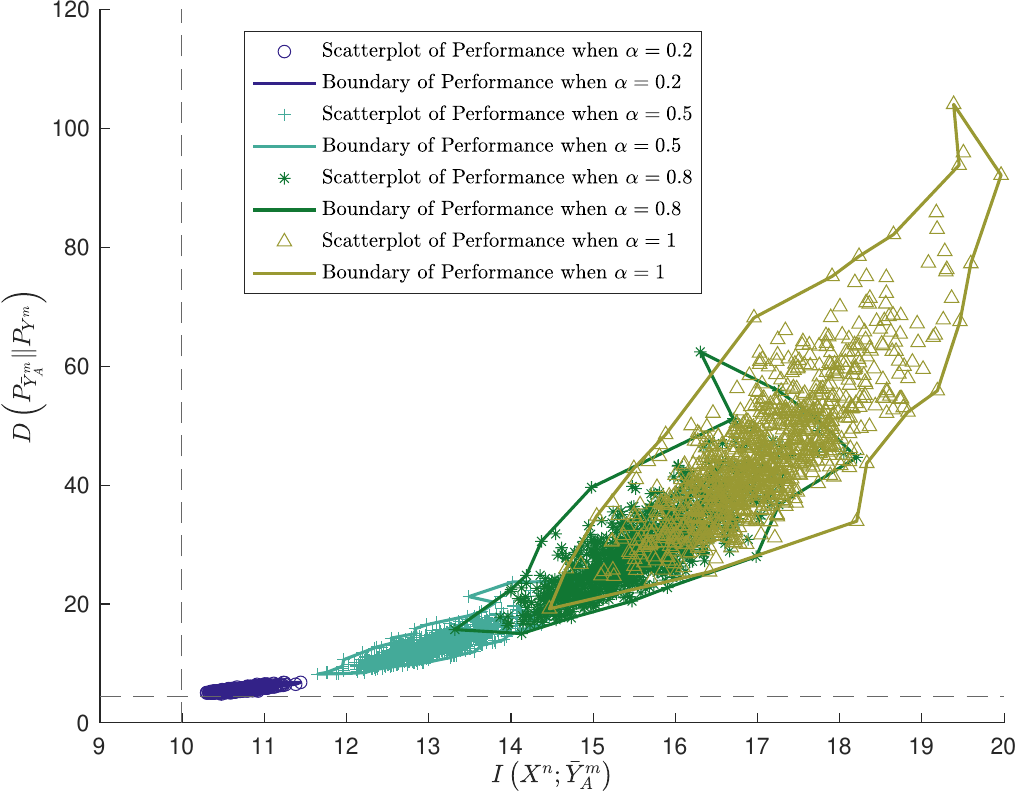}
\caption{Performance of Algorithm \ref{Algo:greedy_1} on IEEE 30-Bus test system when $\textnormal{SNR}=30 \ \textnormal{dB}$ and $\rho=0.5$ under different values of $\alpha$, in which the dashed lines denote the attack performance under the complete admittance information setting.}
\label{Fig:MIKLEnergy_30_SNR30_rho05}
\end{figure}

\begin{figure}[t!]
\centering
\includegraphics[scale=0.48]{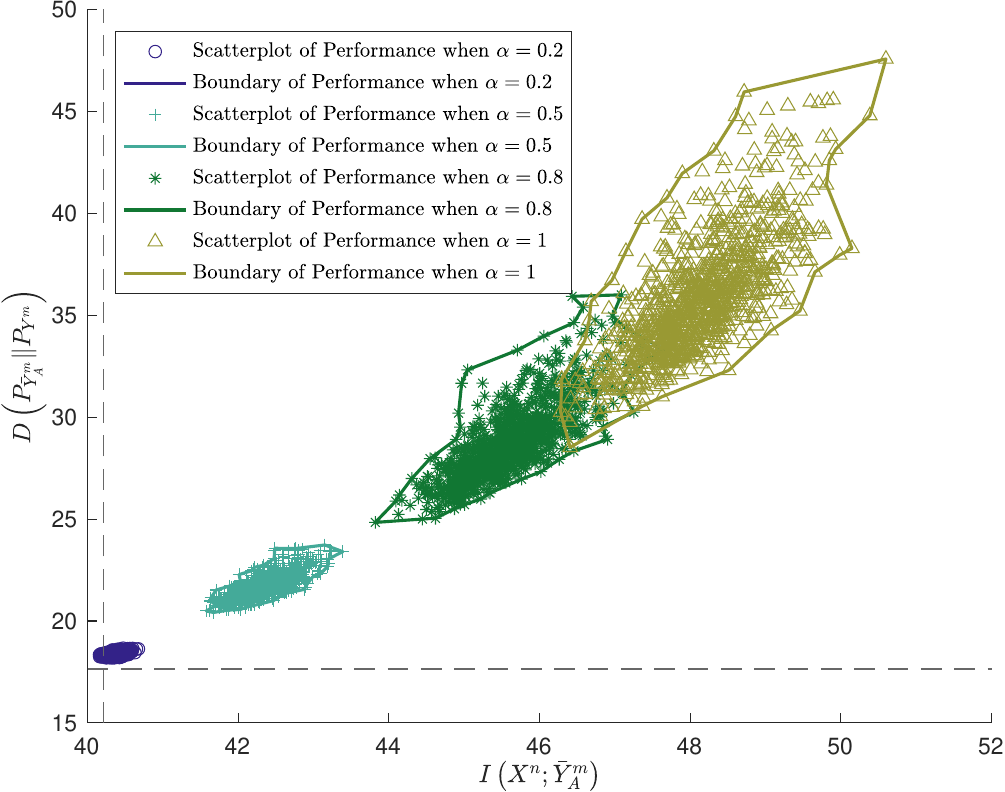}
\caption{Performance of Algorithm \ref{Algo:greedy_1} on IEEE 118-Bus test system when $\textnormal{SNR}=30 \ \textnormal{dB}$ and $\rho=0.5$ under different values of $\alpha$, in which the dashed lines denote the attack performance under the complete admittance information setting.}
\label{Fig:MIKLEnergy_118_SNR30_rho05}
\end{figure}

Fig. \ref{Fig:MIKLEnergy_30_SNR30_rho05} and Fig. \ref{Fig:MIKLEnergy_118_SNR30_rho05} show the performance of Algorithm \ref{Algo:greedy_1}  on IEEE 30-Bus and 118-Bus test system across different values of $\alpha$ when $\textnormal{SNR}=30 \ \textnormal{dB}$ and $\rho=0.5$, in which the dashed lines denote the attack performance under the complete admittance information setting. 
In the simulation, we set $k$ equal to $l$, i.e. all branch admittances are incomplete. 
A total of 1,000 realizations of $\Phim^{max}$ and $\Phim^{min}$ are generated under the constraint that $\phi_{i, max}\ \textnormal{and} \ \phi_{i, max} \in \left[-1, 1\right]$.
This setting ensures that the incomplete and complete branch admittances, i.e. $b_i'$ and $b_i$ in \eqref{Equ:DBAMTD1}, share the same sign and satisfy $0 \leq |b_i'| \leq 2|b_i|$, which aligns with practical constraints faced by system operators.

It can be observed that as the value of $\alpha$ increases, the admissible range for branch admittance becomes wider, allowing Algorithm~\ref{Algo:greedy_1} to select admittance incompleteness values that result in a higher probability of detection. 
Surprisingly, the information obtained by the operator also increases with larger $\alpha$.
This indicates that the enhanced detectability almost does not come at the cost of reduced information gain, which is a ``win-win'' situation for the operator. 
In other words, the incompleteness parameters selected by Algorithm~\ref{Algo:greedy_1} simultaneously both increase the detection probability and reduce the disruptive impact of the attack.

\begin{figure}[t!]
\centering
\includegraphics[scale=0.48]{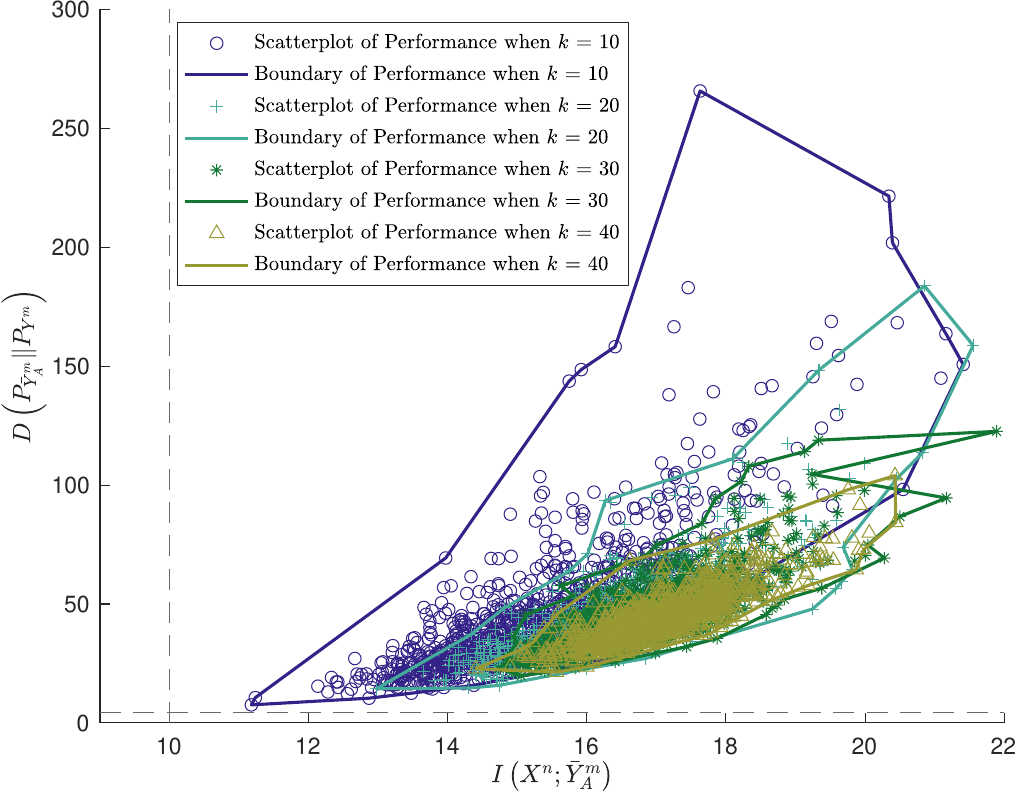}
\caption{Performance of Algorithm \ref{Algo:greedy_1} on IEEE 30-Bus test system when $\textnormal{SNR}=30 \ \textnormal{dB}$, $\rho=0.5$, and $\alpha = 1$ under different values of $k$, in which the dashed lines denote the attack performance under the complete admittance information setting.}
\label{Fig:MIKLkIncrease_30_SNR30_rho05}
\end{figure}

\begin{figure}[t!]
\centering
\includegraphics[scale=0.48]{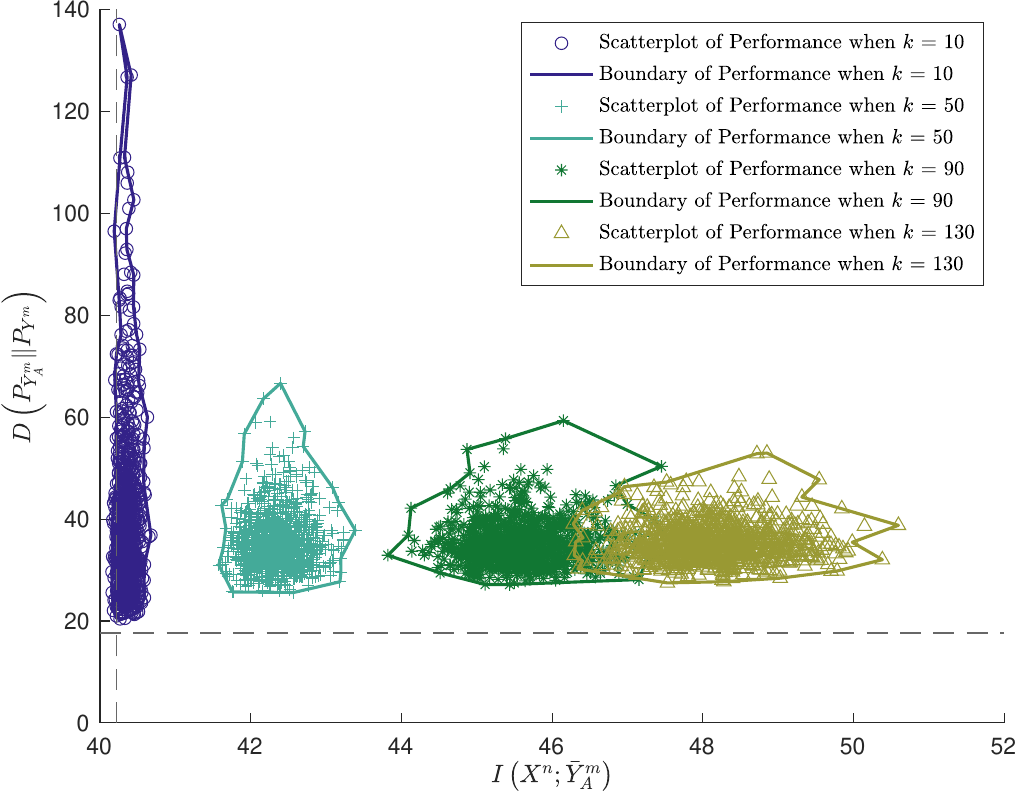}
\caption{Performance of Algorithm \ref{Algo:greedy_1} on IEEE 118-Bus test system when $\textnormal{SNR}=30 \ \textnormal{dB}$, $\rho=0.5$, and $\alpha = 1$ under different values of $k$, in which the dashed lines denote the attack performance under the complete admittance information setting.}
\label{Fig:MIKLkIncrease_118_SNR30_rho05}
\end{figure}

Fig. \ref{Fig:MIKLkIncrease_30_SNR30_rho05} and Fig. \ref{Fig:MIKLkIncrease_118_SNR30_rho05} show the performance of Algorithm \ref{Algo:greedy_1} on the IEEE 30-Bus and 118-Bus test system under varying values of $k$, where $k$ denotes the number of branches whose admittances are incomplete.
The simulation is conducted with $\textnormal{SNR} = 30~\textnormal{dB}$, correlation parameter $\rho = 0.5$, and incompleteness index $\alpha = 1$. 
The dashed lines represent the baseline attack performance under the complete admittance information scenario. 
For each value of $k$, $k$ branches are randomly selected, and their corresponding admittance incompleteness bounds $\phi_{i, min}$ and $\phi_{i, max}$ are independently drawn from the interval $\left[-1, 1\right]$. 
This process is repeated 1,000 times to ensure statistical robustness.

It can be observed that as \(k\) increases, the performance region generally shifts toward lower information loss,  while the distribution of the probability-of-detection metric (i.e. the KL divergence) becomes more concentrated. 
This suggests that incomplete admittance information on a larger number of branches can lead to better information preservation and more consistent detection behavior. 
Moreover, the IEEE 118-Bus system exhibits more compact clustering than the IEEE 30-Bus system.

\subsection{Efficiency of Algorithm \ref{Algo:greedy_1}} 

Algorithm \ref{Algo:greedy_1}  is a greedy algorithm and does not admit a uniform theoretical performance guarantee. 
To assess its practical effectiveness, we evaluate the performance of Algorithm 1 against the optimal solution of optimization problem $\textnormal{\bf P1}$.  
As stated below Theorem \ref{theorem:PDMax}, the cardinality of the reduced candidate solution set is $2^k$. 
Accordingly, we adopt the 9-bus test system from MATPOWER \cite{zimmerman_matpower:_2011} as a benchmark, where the number of branches is $k=9$, yielding a reduced solution set of size $2^9=512$. 

Table \ref{Tab:compare} compares the performance of Algorithm \ref{Algo:greedy_1} with that of the globally optimal solution obtained via brute-force search over the reduced set.
In this evaluation, $\alpha = 1$ and all other settings are identical to those used in Fig. \ref{Fig:MIKLEnergy_30_SNR30_rho05}.
Specifically, we define the performance metric $\beta$ as 
\begin{IEEEeqnarray}{c}
\beta \eqdef \frac{D_{alg}}{D_{bf}},
\end{IEEEeqnarray}
in which $D_{bf}$ and $D_{alg}$ denote the KL divergence (representing probability of detection) induced by the brute-force solution and the solution generated by Algorithm \ref{Algo:greedy_1}, respectively. 
Since $D_{bf}$ corresponds to the optimal solution and the KL divergence is nonnegative, it follows that $\beta \in [0,1]$, with larger values of $\beta$ indicating better performance of Algorithm \ref{Algo:greedy_1}.

The results show that, among 1,000 trials, Algorithm \ref{Algo:greedy_1} attains the global optimum in 338 cases. 
In 228 cases, the performance loss exceeds 10\% (i.e., \(\beta < 0.9\)), whereas only 78 cases exhibit a performance loss greater than 20\% (i.e., \(\beta < 0.8\)).  
For instances with substantial performance degradation (\(\beta < 0.8\)), we observe that the greedy solution often deviates from the brute-force optimum during the assignment of incompleteness values to the first two selected branches, especially the second branch. 
Specifically, only about 35\% of the incompleteness values selected by Algorithm \ref{Algo:greedy_1} for the second branch coincide with the brute-force optimal choices. 
This suggests that severe performance degradation is primarily caused by early-stage greedy decisions. 
Since the algorithm fixes each branch perturbation once it is selected, an early suboptimal assignment can constrain the subsequent search trajectory and prevent the algorithm from reaching the globally optimal vertex.

Ultimately, while Algorithm \ref{Algo:greedy_1} does not guarantee optimality, it yields near-optimal results in most scenarios and rarely suffers from severe performance drops. 
Moreover, for the same pair of $\Phim^{max}$ and $\Phim^{min}$, running Algorithm \ref{Algo:greedy_1} twice and selecting the better solution reduces the probability of a performance loss exceeding 20\% to 0.6\% (i.e. $(78/1000)^2$).

\begin{table}[t!]
\renewcommand{\arraystretch}{1.3}
\caption{values of $\beta$}
\label{Tab:compare}
\centering
\begin{tabular}{c|c|c|c|c|c}
\hline
\bfseries Range of $\beta$ &  1  & [0.95, 1) & [0.9, 0.95) & [0.8, 0.9) & [0, 0.8)\\
\hline
\bfseries Num. of Cases  &338 &304 &130  & 150 & 78  \\
\hline
\end{tabular}
\end{table}

\section{Conclusion}\label{Sec:Con}

Under the Bayesian framework, DIAs need to balance two competing objectives: stealthiness and destructiveness, in which the optimal tradeoff between these objectives depends on the completeness of system information available to the attacker. 
To that end, incomplete system information causes deviations from this optimal tradeoff, thereby degrading the effectiveness of the attack.
In this paper, we focus on the information-theoretic attacks and investigate the impact of the incomplete admittance information on the attack performance degradation. 
Specifically, we establish theoretical sufficient conditions for two distinct operational regimes: (i) stealthiness intensifies while destructive potential diminishes and (ii) destructiveness increases while stealth capability weakens.
For scenarios beyond these regimes, we propose a maximal incompleteness strategy to maximize the degradation in stealthiness. 
The simulation results show that the proposed stealthiness degradation maximization algorithm leads to a third regime, in which (iii) both stealthiness and destructive potential weaken.

\bibliographystyle{IEEEtran}
\bibliography{reference}

\appendices

\section{Proof of Lemma \ref{Lemma:Mismatch}}\label{Sec:ProofMismatch}

Note that from \eqref{Equ:NH}, it holds that 
\begin{IEEEeqnarray}{c}
\Hm' = \Jm \left(\Id + \Phim\right) \Dm \Am = \Jm  \Dm \left(\Id + \Phim\right) \Am,
\end{IEEEeqnarray}
where the last equality follows from the fact that both $\Dm$ in \eqref{Equ:D} and 
$\Phim$ in \eqref{Equ:alpha} are  diagonal matrix, and 
\begin{IEEEeqnarray}{c}
\left(\Id + \Phim\right) \Dm = \Dm \left(\Id + \Phim\right).
\end{IEEEeqnarray}
As a result, it holds that 
\begin{IEEEeqnarray}{c}\label{Equ:ProofMismatch1}
\Hm' \Sigmam_{X\!X} \Hm'^{\sf T} = \Jm  \Dm \left(\Id + \Phim\right) \Am\Sigmam_{X\!X} \Am^{\sf T}\left(\Id + \Phim\right)^{\sf T} \Dm^{\sf T} \Jm^{\sf T}.\IEEEeqnarraynumspace
\end{IEEEeqnarray}
Similarly, from \eqref{Equ:H}, it holds that 
\begin{IEEEeqnarray}{c}\label{Equ:ProofMismatch2}
\Hm \Sigmam_{X\!X} \Hm^{\sf T} = \Jm  \Dm \Am\Sigmam_{X\!X}  \Am^{\sf T}\Dm^{\sf T} \Jm^{\sf T}.
\end{IEEEeqnarray}

Hence, from \eqref{Equ:ProofMismatch1} and \eqref{Equ:ProofMismatch2}, it holds that 
\begin{IEEEeqnarray}{rl}
&\Hm' \Sigmam_{X\!X} \Hm'^{\sf T} - \Hm \Sigmam_{X\!X} \Hm^{\sf T} \nonumber \\
& = \Jm \Dm \left( \left(\Id + \Phim\right) \Am\Sigmam_{X\!X} \Am^{\sf T}\left(\Id + \Phim\right)^{\sf T} - \Am\Sigmam_{X\!X}  \Am^{\sf T}\right) \Dm^{\sf T} \Jm^{\sf T} \squeezeequ \IEEEeqnarraynumspace \\
& =\Jm \Dm  \Deltam\Dm^{\sf T} \Jm^{\sf T}, \label{Equ:ProofMismatch3} \squeezeequ \IEEEeqnarraynumspace
\end{IEEEeqnarray}
And combining \eqref{Equ:ProofMismatch2} and \eqref{Equ:ProofMismatch3} yields the equality in \eqref{Equ:Mismatch1}. 

This completes the proof.

\section{Proof of Lemma \ref{Lemma:PSD}}\label{Sec:ProofPSD}

From \eqref{Equ:Delta}, it is easy to find that $\Am\Sigmam_{X\!X}  \Am^{\sf T} +\Deltam$ is a symmetric matrix. 

We use the proof-by-contradiction method to prove this lemma. 
Suppose that $\Am\Sigmam_{X\!X}  \Am^{\sf T} +\Deltam$ is not a PSD matrix. 
Hence, $\exists \bv \in \Rc^{l}/{\zerov}$, it holds that 
\begin{IEEEeqnarray}{c}
\bv^{\sf T} \left(\Am\Sigmam_{X\!X}  \Am^{\sf T} +\Deltam \right) \bv < 0. 
\end{IEEEeqnarray}
Let $\Mm \eqdef \Jm \Dm \left( (\Jm \Dm)^{\sf T} \Jm \Dm\right)^{-1}$. 
Given the fact that $\Jm$ and $\Dm$ are both full-rank matrices, it holds that $\Mm$ is a full rank matrix. 
As a result, it holds for vector $ \Mm \bv \in \Rc^{m}/\zerov$ that 
\begin{IEEEeqnarray}{rl}
    &(\Mm \bv)^{\sf T} \Jm \Dm \left(\Am\Sigmam_{X\!X}  \Am^{\sf T} +\Deltam \right) \Dm^{\sf T} \Jm^{\sf T}\Mm \bv \\
    & = \bv^{\sf T} \Mm^{\sf T}\Jm \Dm \left(\Am\Sigmam_{X\!X}  \Am^{\sf T} +\Deltam \right) \Dm^{\sf T} \Jm^{\sf T}\Mm \bv \\
    & = \bv^{\sf T} \left(\Am\Sigmam_{X\!X}  \Am^{\sf T} +\Deltam \right) \bv \label{Equ:LemmaPSD1} \\
    &< 0,
\end{IEEEeqnarray}
where \eqref{Equ:LemmaPSD1} follows from the fact that 
\begin{IEEEeqnarray}{c}
\Mm^{\sf T}\Jm \Dm = \left(\left( (\Jm \Dm)^{\sf T} \Jm \Dm\right)^{-1}\right)^{\sf T} (\Jm \Dm)^{\sf T} \Jm \Dm = \Id,
\end{IEEEeqnarray}
in which $(\Jm \Dm)^{\sf T} \Jm \Dm$ is a symmetric matrix. 
This implies that $\Jm \Dm \left(\Am\Sigmam_{X\!X}  \Am^{\sf T} +\Deltam \right) \Dm^{\sf T} \Jm^{\sf T}$ is not a PSD matrix. 
However, from Lemma \ref{Lemma:Mismatch}, it is proved that $\Jm \Dm \left(\Am\Sigmam_{X\!X}  \Am^{\sf T} +\Deltam \right) \Dm^{\sf T} \Jm^{\sf T}$ is a PSD matrix, since it is the covariance matrix of $\bar{A}^m$.
To that end, it holds that $\Am\Sigmam_{X\!X}  \Am^{\sf T} +\Deltam$ is a PSD matrix.

This completes the proof.

\section{Proof of Lemma \ref{Lemma:YBarADis}}\label{Sec:ProofYBarADis}
Under the setting that $X^n$, $Z^{m}$, and $\bar{A}^m$ are all zero-mean multivariate Gaussian random variables, from \eqref{Equ:YBarA}, it is known that  $\bar{Y}_{A}^m $ is also zero-mean multivariate Gaussian random variables. 

Furthermore, for the covariance matrix of $\bar{Y}_{A}^m $, it holds that 
\begin{IEEEeqnarray}{rl}
\Sigmam_{\bar{Y}_A\bar{Y}_A} &= \Sigmam_{YY} + \Hm' \Sigmam_{X\!X}\Hm'^{\sf T} \\
& = \Jm \Dm \Am\Sigmam_{X\!X} \Am^{\sf T} \Dm^{\sf T} \Jm^{\sf T} +\sigma^2\Id \IEEEnonumber \\
& \qquad + \Jm  \Dm \left(\Am\Sigmam_{X\!X}  \Am^{\sf T} +\Deltam\right)\Dm^{\sf T} \Jm^{\sf T} \label{Equ::ProofYBarADis1} \\
& = \Jm \Dm \left(2\Am\Sigmam_{X\!X} \Am^{\sf T} + \Deltam \right)\Dm^{\sf T} \Jm^{\sf T}+\sigma^2\Id \IEEEeqnarraynumspace
\end{IEEEeqnarray}
where \eqref{Equ::ProofYBarADis1} follows from \eqref{Equ:YMDis} and \eqref{Equ:Mismatch2}. 

This completes the proof.

\section{Proof of Lemma \ref{Lemma:PSD-G}}\label{Sec:ProofPSD-G}

Note that if $\Xm$ is a PSD matrix, it holds that $\Xm$ is a symmetric matrix. 
As a result, it holds that $ \Mm \Xm \Mm^{\sf T}$ is a symmetric matrix.

In the following, we prove that the minimum eigenvalue of $\Mm \Xm \Mm^{\sf T}$ is greater than or equal to zero. 
Let $\lambda_{\textnormal{min}}(\Xm)$ denote the minimum eigenvalue of matrix $\Xm$.
Given the fact that $\Mm \Xm \Mm^{\sf T}$ and $\Xm  \Mm^{\sf T}\Mm$ have the same nonzero eigenvalues \cite[6.54]{seber_matrix_2008} and are of the same dimensions, 
it holds that 
\begin{IEEEeqnarray}{rl}
\lambda_{\textnormal{min}} \left(\Mm \Xm \Mm^{\sf T}\right) & = \lambda_{\textnormal{min}} \left(\Xm \Mm^{\sf T}\Mm \right) \label{Equ:lemmaPD1}\\
& \geq \lambda_{\textnormal{min}} \left(\Xm \right)\lambda_{\textnormal{min}} \left(\Mm^{\sf T}\Mm  \right) \label{Equ:lemmaPD2} \\
&\geq 0 \label{Equ:lemmaPD3} 
\end{IEEEeqnarray}
where  \eqref{Equ:lemmaPD2} follows from the fact that  $\Mm^{\sf T}\Mm $ is a PSD matrix and \cite[6.76]{seber_matrix_2008}; 
and \eqref{Equ:lemmaPD3} follows from the fact that both $\Xm$ and $\Mm^{\sf T}\Mm $ are PSD matrices. 

This completes the proof.

\section{Proof of Lemma \ref{Lemma:logdet_mono}}\label{Sec:ProoflogdetMono}

Note that from Lemma \ref{Lemma:PSD-G}, it holds that $\Mm\Xm^{-1}\Mm^{\sf T}$ is a PSD matrix. 
Hence, the function $\log \left| \Id + \Mm\Xm^{-1}\Mm^{\sf T}\right|$ is well-defined for all $\Mm \in \Rc^{m \times m}$ and $\Xm \in \Sc^{m}_{++}$.

Let $\Xm_1, \Xm_2 \in \Sc_{++}^{m}$ such that 
\begin{IEEEeqnarray}{c}
\Xm_1- \Xm_2 = \Delta\Xm \succ \zerov. \label{Equ:Lemmamono1}
\end{IEEEeqnarray}
Hence, it follows from \cite[10.51(a)]{seber_matrix_2008} that 
\begin{IEEEeqnarray}{c}
\Xm_1^{-1} \prec \Xm_2^{-1}.
\end{IEEEeqnarray}
To that end, from \cite[Lemma V.1.5]{bhatia_2013_matrix}, it holds that 
\begin{IEEEeqnarray}{c}
\Id + \Mm\Xm_1^{-1}\Mm^{\sf T} \preceq \Id +\Mm\Xm_2^{-1}\Mm^{\sf T}.
\end{IEEEeqnarray}
Hence, from \cite[Example 3.46]{boyd_convex_2004}, it holds that 
\begin{IEEEeqnarray}{c}
\log \left| \Id + \Mm\Xm_1^{-1}\Mm^{\sf T}\right| \leq \log \left| \Id + \Mm\Xm_2^{-1}\Mm^{\sf T}\right|,
\end{IEEEeqnarray}
which implies that $\log \left| \Id + \Mm\Xm_1^{-1}\Mm^{\sf T}\right|$ is a matrix-nonincreasing function of $\Xm$.

This completes the proof. 

\section{Proof of Lemma \ref{Lemma:CND}}\label{Sec:ProofCND}

From \cite[pp. 73]{boyd_convex_2004}, it is known that $-\log \left| \Id + \Xm\right|$ is a convex function of $\Xm$.
Adding the fact that the trace operator is a linear operator, it is known that function $-\log |\Id + \Xm| + \trace(\Xm)$ is a convex function of $\Xm$ on $\Sc_{+}^{m}$. 

Now we proceed to prove the matrix-monotonic of the function. 
Let $\lambda_{i}\left( \Xm\right)$, $\lambda_{\textnormal{min}}\left( \Xm\right)$, and $\lambda_{\textnormal{max}}\left( \Xm\right)$ denote the $i$-th, the minimum, and the maximimum eigenvalue of $\Xm$, respectively. 
And let $\Xm_1, \Xm_2 \in \Sc_{+}^{m}$ such that 
\begin{IEEEeqnarray}{c}
\Xm_1- \Xm_2 = \Delta\Xm \succ \zerov. \label{Equ:LemmaCND1}
\end{IEEEeqnarray}
Hence, it holds that 
\begin{IEEEeqnarray}{l}
-\log |\Id+ \Xm_1| + \trace(\Xm_1) + \log |\Id+\Xm_2| - \trace(\Xm_2) \nonumber \\
\ = \trace(\Delta\Xm) - \log |(\Id + \Xm_2)^{-1}(\Id + \Xm_2 + \Delta \Xm)| \label{Equ:LemmaCND2}\\
\ = \trace(\Delta\Xm) - \log |\Id + (\Id + \Xm_2)^{-1} \Delta \Xm| \\
\ \geq \sum_{i=1}^m \lambda_i (\Delta\Xm) - \sum_{i=1}^m \log\left( 1 + \frac{1}{1+\lambda_{\textnormal{min}}(\Xm_2)} \lambda_i (\Delta\Xm)\right) \label{Equ:LemmaCND3}\IEEEeqnarraynumspace \squeezeequ\\
\ \geq \sum_{i=1}^m \Bigl(\lambda_{i}(\Delta \Xm) - \log \left( 1+ \lambda_{i}\left(\Delta \Xm\right)\right) \Bigr)\label{Equ:LemmaCND4}\\
\ \geq 0\label{Equ:LemmaCND5}
\end{IEEEeqnarray}
where \eqref{Equ:LemmaCND2} follows from \eqref{Equ:LemmaCND1};
\eqref{Equ:LemmaCND3} follows from the fact that 
\begin{IEEEeqnarray}{rl}
\lambda_i \left((\Id + \Xm_2)^{-1} \Delta \Xm\right) &\leq \lambda_{\textnormal{max}} \left((\Id + \Xm_2)^{-1}\right) \lambda_i (\Delta \Xm) \IEEEeqnarraynumspace\\
& \leq \frac{1}{1+\lambda_{\textnormal{min}} (\Xm_2)}\lambda_i (\Delta \Xm),
\end{IEEEeqnarray}
which is a result of \cite[6.76]{seber_matrix_2008}; 
\eqref{Equ:LemmaCND4} follows from the fact that $\lambda_{\textnormal{min}}(\Xm_2) \geq 0$; 
and \eqref{Equ:LemmaCND5} follows from the fact that $\lambda_{i}\left(\Delta \Xm\right) \geq 0$, and when $x \geq 0$,
\begin{IEEEeqnarray}{c}
f(x) = x -\log (1+x)
\end{IEEEeqnarray}
is a monotonically increasing function, whose minimum value $0$ is attained at $x=0$.
This implies that function $-\log |\Id + \Xm| + \trace(\Xm)$ is a matrix-nondecreasing function of $\Xm$.

Now we proceed to prove the minimum value of the function.
It holds for all $\mathbf{X}\in \mathcal{S}_{+}^{m}$ that 
\begin{IEEEeqnarray}{c}
\mathbf{X}\succeq \zerov. 
\end{IEEEeqnarray}
Since $-\log |\Id + \Xm| + \trace(\Xm)$ is a nondecreasing function of $\Xm$, it holds for all $\mathbf{X}\in \mathcal{S}_{+}^{m}$ that 
\begin{IEEEeqnarray}{c}
-\log |\Id + \Xm| + \trace(\Xm) \geq -\log |\Id + \zerov| + \trace(\zerov) = 0. \IEEEeqnarraynumspace
\end{IEEEeqnarray}

This completes the proof.

\section{Proof of Theorem \ref{Lemma:Delta_Con}}\label{Sec:ProofDeltaCon}

Note that if $\Deltam \succeq \zerov$, it holds that 
\begin{IEEEeqnarray}{c}\label{Equ:PSD-1}
\Am \Sigmam_{X\!X} \Am^{\sf T} + \Deltam \succeq \Am \Sigmam_{X\!X} \Am^{\sf T}. 
\end{IEEEeqnarray}
Then from \cite[Lemma V.1.5]{bhatia_2013_matrix} , it holds that 
\begin{IEEEeqnarray}{c}
\Jm \Dm \left(\Am \Sigmam_{X\!X} \Am^{\sf T} + \Deltam \right) \Dm^{\sf T} \Jm^{\sf T}\succeq \Jm \Dm \Am \Sigmam_{X\!X} \Am^{\sf T}\Dm^{\sf T} \Jm^{\sf T}. \label{Equ:Delta_Con1}  \IEEEeqnarraynumspace \squeezeequ
\end{IEEEeqnarray}

Firstly, we prove the statement about $D\left(P_{\bar{Y}_{A}^m} \| P_{Y^m}\right) $. 
For \eqref{Equ:Delta_Con1}, reusing the result in \cite[Lemma V.1.5]{bhatia_2013_matrix} yields that 
\begin{IEEEeqnarray}{rl}
&\Sm^{\frac{1}{2}}\Jm \Dm \left(\Am \Sigmam_{X\!X} \Am^{\sf T} + \Deltam \right) \Dm^{\sf T} \Jm^{\sf T} \Sm^{\frac{1}{2}} \IEEEnonumber\\
& \qquad \qquad \qquad \qquad \succeq \Sm^{\frac{1}{2}}\Jm \Dm \Am \Sigmam_{X\!X} \Am^{\sf T}\Dm^{\sf T} \Jm^{\sf T}\Sm^{\frac{1}{2}}. \IEEEeqnarraynumspace \squeezeequ
\end{IEEEeqnarray}
Hence, from \eqref{Equ:KL} and Lemma \ref{Lemma:CND}, it holds that 
\begin{IEEEeqnarray}{c}
D\left(P_{\bar{Y}_{A}^m} \| P_{Y^m}\right)  \geq D\left(P_{\tilde{Y}_A^m} \| P_{Y^m}\right) 
\end{IEEEeqnarray}

Then we prove the statements about $I\left(X^{n};\bar{Y}^{m}_{A}\right)$. 
If $\Deltam \succeq \zerov$, it holds that 
\begin{IEEEeqnarray}{c}
\Am \Sigmam_{X\!X} \Am^{\sf T} + \Deltam \succeq \Am \Sigmam_{X\!X} \Am^{\sf T}. 
\end{IEEEeqnarray}
Then from \cite[Lemma V.1.5]{bhatia_2013_matrix}, it holds that 
\begin{IEEEeqnarray}{c}
\Jm \Dm \left(\Am \Sigmam_{X\!X} \Am^{\sf T} + \Deltam \right) \Dm^{\sf T} \Jm^{\sf T}\succeq \Jm \Dm \Am \Sigmam_{X\!X} \Am^{\sf T}\Dm^{\sf T} \Jm^{\sf T}.  \IEEEeqnarraynumspace \squeezeequ
\end{IEEEeqnarray}
Hence, from \eqref{Equ:DeltaCon1} and Lemma \ref{Lemma:logdet_mono}, it holds that 
\begin{IEEEeqnarray}{c}
I\left(X^{n};\bar{Y}^{m}_{A}\right) \leq I\left(X^{n};\tilde{Y}^{m}_{A}\right).
\end{IEEEeqnarray}

The proof completes by noticing that the case $\Deltam \preceq \zerov$ follows the same way.

\section{Proof of Lemma \ref{Lemma:SC_Delta}}\label{Sec:ProofDeltaSC}
For \eqref{Equ:Delta}, it holds that 
\begin{IEEEeqnarray}{c}
\Phim \Am \Sigmam_{X\!X} \Am^{\sf T} = (\phiv\onev^{\sf T})\odot \Am \Sigmam_{X\!X} \Am^{\sf T} \\
 \Am \Sigmam_{X\!X} \Am^{\sf T}\Phim^{\sf T} = (\onev\phiv^{\sf T})\odot \Am \Sigmam_{X\!X} \Am^{\sf T} \\
\Phim \Am \Sigmam_{X\!X} \Am^{\sf T} \Phim^{\sf T } = (\phiv\phiv^{\sf T})\odot \Am \Sigmam_{X\!X} \Am^{\sf T},
\end{IEEEeqnarray}
where $\odot$ is the Hadamard product. 
To that end, it holds for $\Deltam$ in \eqref{Equ:Delta} that 
\begin{IEEEeqnarray}{c}\label{Equ:last1}
\Deltam = (\phiv\phiv^{\sf T}+\phiv\onev^{\sf T}+\onev\phiv^{\sf T})\odot \Am \Sigmam_{X\!X} \Am^{\sf T} = \tilde{\Phim} \odot \Am \Sigmam_{X\!X} \Am^{\sf T},  \supersqueezeequ \IEEEeqnarraynumspace
\end{IEEEeqnarray}
where $\tilde{\Phim} \eqdef \phiv\phiv^{\sf T}+\phiv\onev^{\sf T}+\onev\phiv^{\sf T}$.

Given the fact that $\phiv\onev^{\sf T}$ and $\onev\phiv^{\sf T}$ in $\tilde{\Phim}$ are all rank one matrix, it holds that symmetric matrix $\phiv\onev^{\sf T}+\onev\phiv^{\sf T}$ is at most with rank $2$ and has at most two nonzero real eigenvalues. Let $\lambda_1$ and $\lambda_2$ denote the eigenvalues of $\phiv\onev^{\sf T}+\onev\phiv^{\sf T}$. 
It holds that 
\begin{IEEEeqnarray}{c}
\lambda_1 + \lambda_2 = \trace\left(\phiv\onev^{\sf T}+\onev\phiv^{\sf T}\right) = \phiv^{\sf T} \onev + \onev^{\sf T} \phiv = 2  \phiv^{\sf T} \onev \IEEEeqnarraynumspace
\end{IEEEeqnarray}
and 
\begin{IEEEeqnarray}{c}
\lambda_1^2 + \lambda_2^2 = \trace\left(\left(\phiv\onev^{\sf T}+\onev\phiv^{\sf T}\right)^2\right) = 2 \left(\left(\phiv^{\sf T} \onev \right)^2 + \phiv^{\sf T}\phiv \onev^{\sf T} \onev \right), \Tsupersqueezeequ \IEEEeqnarraynumspace
\end{IEEEeqnarray}
from which it holds that 
\begin{IEEEeqnarray}{c}
\lambda_1  \lambda_2 = \frac{1}{2}\left(\left(\lambda_1 + \lambda_2\right)^2 - \lambda_1^2 - \lambda_2^2 \right)= \left(\phiv^{\sf T} \onev \right)^2 - \phiv^{\sf T}\phiv \onev^{\sf T} \onev. \squeezeequ \IEEEeqnarraynumspace
\end{IEEEeqnarray}
Hence, $\lambda_1$ and $\lambda_2$ are solutions to the quadratic function
\begin{IEEEeqnarray}{c}
\lambda^2 - 2\left(\phiv^{\sf T} \onev\right)\lambda  + \left(\phiv^{\sf T} \onev \right)^2 - \phiv^{\sf T}\phiv \onev^{\sf T} \onev = 0, \label{Equ:h2} \IEEEeqnarraynumspace
\end{IEEEeqnarray}
from which it holds that 
\begin{IEEEeqnarray}{c}
\lambda_1 = \phiv^{\sf T} \onev + \sqrt{\phiv^{\sf T}\phiv \onev^{\sf T} \onev}, \quad 
\lambda_2 = \phiv^{\sf T} \onev - \sqrt{\phiv^{\sf T}\phiv \onev^{\sf T} \onev}, \IEEEeqnarraynumspace
\end{IEEEeqnarray}
and $\lambda_1 \geq \lambda_2$. 

Note that $\phiv\phiv^{\sf T}$ in $\tilde{\Phim}$ is a rank one symmetric matrix with only one nonzero eigenvalue, whose value equals to $\phiv^{\sf T}\phiv \geq 0$.
To that end, it follows from \cite[6.71]{seber_matrix_2008} that 
\begin{IEEEeqnarray}{rl}
\lambda_{max}(\tilde{\Phim}) &\leq \lambda_{max}(\phiv\phiv^{\sf T}) + \lambda_{max}(\phiv\onev^{\sf T}+\onev\phiv^{\sf T}) \\
& =\phiv^{\sf T} \phiv +  \phiv^{\sf T} \onev + \sqrt{\phiv^{\sf T}\phiv \onev^{\sf T} \onev}
\end{IEEEeqnarray}
and 
\begin{IEEEeqnarray}{rl}
\lambda_{min}(\tilde{\Phim}) &\geq \lambda_{min}(\phiv\phiv^{\sf T}) + \lambda_{min}(\phiv\onev^{\sf T}+\onev\phiv^{\sf T}) \\
& = \phiv^{\sf T} \onev - \sqrt{\phiv^{\sf T}\phiv \onev^{\sf T} \onev}.
\end{IEEEeqnarray}

To that end, if the condition in \eqref{Equ:Mon1} holds, it holds that $\lambda_{min}(\tilde{\Phim}) \geq 0$, which suggests that $\tilde{\Phim}$ is a PSD matrix. 
And from \eqref{Equ:last1} and \cite[11.46]{seber_matrix_2008}, it holds that $\Deltam \succeq \zerov$. 
Similarly, if the condition in \eqref{Equ:Mon2} holds, it holds that 
$\lambda_{max}(\tilde{\Phim}) \leq 0$, which suggests that $-\tilde{\Phim}$ is a PSD matrix, and $\Deltam \preceq \zerov$ \cite[11.46]{seber_matrix_2008}. 

This completes the proof. 

\section{Proof of Theorem \ref{theorem:PDMax}}\label{Sec:ProofPDMax}

The convexity of the objective function is proved first. 
From Lemma \ref{Lemma:CND}, it is known that $-\log \left| \Id + \Sm^{\frac{1}{2}} \Tm \Sm^{\frac{1}{2}} \right| + \trace \left( \Sm^{\frac{1}{2}} \Tm \Sm^{\frac{1}{2}} \right) $ is a convex and  matrix-nondecreasing function of $\Sm^{\frac{1}{2}} \Tm \Sm^{\frac{1}{2}} $.

To that end, to prove the convexity of the objective function, we only need to prove that $\Sm^{\frac{1}{2}} \Tm \Sm^{\frac{1}{2}}$ is a matrix-convex function of $\Phim$ \cite[pp. 84]{boyd_convex_2004}.
Or equivalently, from \eqref{Equ:Tm}, we only need to prove that $\Sm^{\frac{1}{2}} \Jm \Dm \Deltam \Dm^{\sf T} \Jm^{\sf T}\Sm^{\frac{1}{2}}$ is a convex function of $\Phim$.
Let $\Mm \eqdef \Sm^{\frac{1}{2}} \Jm \Dm$. 
With a little abuse of notation, we denote $\Deltam$ as a function of $\Phim$, i.e. $\Deltam(\Phim)$.
It holds for $\theta \in [0,1]$ such that 
\begin{IEEEeqnarray}{l}
\Mm \left(\Deltam(\theta\Phim_1 + (1-\theta)\Phim_2)\right)\Mm^{\sf T} \nonumber  \\
\quad \preceq \Mm \left(\theta\Deltam(\Phim_1) + (1-\theta)\Deltam(\Phim_2)\right)\Mm^{\sf T} \label{Equ:PDMax_1} \\
\quad \preceq \theta \Mm \Deltam(\Phim_1)\Mm^{\sf T} + (1-\theta)\Mm \Deltam(\Phim_2)\Mm^{\sf T}
\end{IEEEeqnarray}
where \eqref{Equ:PDMax_1} follows from the fact that $\Deltam$ is a matrix-convex function of $\Phim$ \cite[Example 3.49]{boyd_convex_2004} and \cite[Lemma V.1.5]{bhatia_2013_matrix}. 
This implies that $\Sm^{\frac{1}{2}} \Jm \Dm \Deltam \Dm^{\sf T} \Jm^{\sf T}\Sm^{\frac{1}{2}}$ is a matrix-convex function of $\Phim$, which completes the proof of the convexity of the objective function.

Adding the fact that constraint $\Phim^{min} \leq \Phim \leq  \Phim^{max}$ forms a closed convex set (more specifically, a polyhedron), it is proved that a convex function of $\Phim$ is maximized within a convex set in problem $\bf P1$.

Furthermore, from \cite{tardella_1988_class}, it is known that maximizing a convex function over a polyhedron is equivalent to maximizing that function over the vertices of the polyhedron. 
To that end, the optimal solution $ \Phim^{\star}$ belongs to the set given by \eqref{Equ:PDmax1}.

This completes the whole proof.

\section{Proof of Lemma \ref{Lemma:Ind}}\label{Sec:ProofInd}
Note that for $\Deltam$ in \eqref{Equ:Delta}, it holds that
\begin{IEEEeqnarray}{rl}
\Deltam & = (\Id+\Phim) \Am\Sigmam_{X\!X}\Am^{\sf T}(\Id+\Phim)^{\sf T} - \Am\Sigmam_{X\!X} \Am^{\sf T} \nonumber\\ 
&= (\Id+\Phim) \Nm(\Id+\Phim)^{\sf T} - \Nm.
\end{IEEEeqnarray}
Let $\ev_i$ and $\ev_j$ be the standard basis vectors, in which only the $i$-th and $j$-th elements are the only nonzero element and equal to 1. 
If the condition in \eqref{Equ:Ind_con1} holds, then 
\begin{IEEEeqnarray}{l}
\ev_i^{\sf T}\Deltam \ev_i = \left((1+\phi_i)^2 - 1\right)N_{ii} = d_i N_{ii} < 0, \\
\ev_j^{\sf T}\Deltam \ev_j = \left((1+\phi_j)^2 - 1\right)N_{jj} = d_j N_{jj} > 0,
\end{IEEEeqnarray}
which  implies that $\Deltam$ is indefinite. 

The case where condition in \eqref{Equ:Ind_con2} holds follows by the same argument. 
This completes the proof.

\end{document}